\documentclass[10pt,conference,letterpaper]{IEEEtran}
\makeatletter
\let\old@citex\@citex
\makeatother
\usepackage[english]{babel}
\makeatletter
\let\@citex\old@citex
\makeatother
\usepackage{mathptmx,helvet}
\usepackage{color}
\usepackage{multirow}
\usepackage{PageSetting}
\usepackage{graphicx}
\usepackage{balance}
\usepackage{times}
\usepackage{algorithm}
\usepackage[noend]{algorithmic}
\usepackage{subfigure}
\usepackage{comment}
\usepackage{hyperref}
\usepackage{extarrows}

\title{Scalable De Novo Genome Assembly Using Pregel}

\author{
Da Yan{\small $^\#$}, Hongzhi Chen{\small $^*$}, James Cheng{\small $^*$}, Zhenkun Cai{\small $^*$}, Bin Shao{\small $^\dagger$}
\vspace{1.6mm}\\
\fontsize{10}{10}\selectfont\itshape
\fontsize{10}{10}\selectfont\itshape\rmfamily $^\#$Department of Computer Science, The University of Alabama at Birmingham\\
\fontsize{9}{9}\selectfont\ttfamily\upshape $^\#$yanda@uab.edu\\
\fontsize{10}{10}\selectfont\itshape\rmfamily $^*$Department of Computer Science and Engineering, The Chinese University of Hong Kong\\
\fontsize{9}{9}\selectfont\ttfamily\upshape $^*$\{hzchen, jcheng, zkcai\}@cse.cuhk.edu.hk\\
\fontsize{10}{10}\selectfont\itshape\rmfamily $^\dagger$Microsoft Research Asia\\
\fontsize{9}{9}\selectfont\ttfamily\upshape $^\dagger$binshao@microsoft.com\\
\vspace{-8mm}
}

\begin{document}
\maketitle

\begin{abstract}
De novo genome assembly is the process of stitching short DNA sequences to generate longer DNA sequences, without using any reference sequence for alignment. It enables high-throughput genome sequencing and thus accelerates the discovery of new genomes. In this paper, we present a toolkit, called PPA-assembler, for de novo genome assembly in a distributed setting. The operations in our toolkit provide strong performance guarantees, and can be assembled to implement various sequencing strategies. PPA-assembler adopts the popular {\em de Bruijn graph} based approach for sequencing, and each operation is implemented as a program in Google's Pregel framework for big graph processing. Experiments on large real and simulated datasets demonstrate that PPA-assembler is much more efficient than the state-of-the-arts and provides good sequencing quality.
\end{abstract}

\section{Introduction}
Modern sequencing technologies generate a very large number of short DNA segments called {\em reads}, which are stitched together to generate longer DNA sequences for finding new genomes. Although millions of reads can be generated in a day to allow high sequencing coverage, the assembly process becomes highly computation-intensive. Single-threaded assemblers often require a high-end server with terabytes of RAM, and are not efficient enough. As a result, parallel short read assembly has aroused a lot of attention recently thanks to the advances in big data systems. Many parallel (and often, distributed) assemblers have emerged, including ABySS~\cite{abyss}, Spaler~\cite{spaler}, Ray~\cite{ray} and SWAP-Assembler~\cite{swap}.

Instead of providing yet another parallel assembler, we developed a toolkit called PPA-assembler (\url{https://github.com/yaobaiwei/PPA-Assembler}), which implements the basic data structures and operations in genome assembly. The operations have strong performance guarantees, and can be assembled to implement various sequencing strategies. Each operation may either read its input from Hadoop Distributed File System (HDFS)\footnote{\scriptsize Hadoop: \url{http://hadoop.apache.org/}}, or directly obtain its input by converting the output of another operation in memory. As a result, PPA-assembler can inter-operate with existing Big Data platforms such as Hadoop and Spark~\cite{spark}, and intermediate results between consecutive jobs do not have to go through HDFS.

PPA-assembler adopts the popular {\em de Bruijn graph} (DBG) based approach for sequencing~\cite{Pevzner14082001}. Therefore, we built it on top of Pregel+\footnote{\scriptsize Pregel+: \url{http://www.cse.cuhk.edu.hk/pregelplus/}}, our open-source implementation of Google's Pregel framework for big graph processing. We remark that our solution is applicable to any Pregel-like system~\cite{giraph,gps,graphx}, and Pregel+ is adopted mainly due to its superior performance as reported by~\cite{yiExp} and due to its wide application~\cite{FengCLQZ16,ppa,da_www}. Since the assembly process also involves some non-graph operations, such as to construct DBG from raw DNA reads, we also extended Pregel+'s API with new functionalities, including grouping and merging data by key, and in-memory data conversion for seamless job concatenation.

We summarized the key operations from existing assemblers, such as contig merging, tip removing and bubble filtering, and implemented these operations in PPA-assembler as scalable Pregel programs. Specifically, each operation is implemented as a {\em Practical Pregel Algorithm} (PPA) as defined in~\cite{ppa}, which runs for at most logarithmic number of iterations (to DBG size), and each iteration has linear space usage, computation cost and communication cost. Users may also include new operations into PPA-assembler by implementing them using the user-friendly Pregel API.

Unlike ABySS~\cite{abyss}, Ray~\cite{ray} and SWAP-Assembler~\cite{swap}, PPA-assembler decouples low-level execution (e.g., data distribution and communication) from the high-level assembly logic, allowing both layers to be independently optimized. For example, as the {\em Implementation} section of~\cite{abyss} indicates, ABySS needs to collect messages into larger 1KB packets for transmission in batch, in order to hide the round-trip time of individual messages, but such communication details are automatically taken care of and optimized by a Pregel-like system. Moreover, PPA-assembler is compatible with existing big data platforms and can inter-operate with other systems that perform various sequence mining and analytics tasks.

Although Spaler~\cite{spaler} is built on top of Spark to be compatible with existing big data platforms, the algorithms designed are rather ad hoc: they only demonstrate how genome assembly operations can be mapped into Spark API, without any formal analysis on the computation complexity. Moreover, most operations in DBG-based sequencing are graph operations, for which Spaler~\cite{spaler} uses the GraphX (Spark's graph API) that are often over one order of magnitude slower than tailer-made Pregel-like systems~\cite{husky,pregelix}. PPA-assembler adopts the efficient Pregel+ system to process the dominating graph operations in DBG-based sequencing, and extends the API to conveniently and efficiently support those non-graph operations required during assembly. We compared PPA-assembler with other existing distributed assemblers on large simulated and real datasets, and the results show that PPA-assembler significantly beats all other assemblers in execution time, and provides comparable sequencing quality.

The rest of this paper is organized as follows. Section~\ref{sec:pregel} reviews the framework of Pregel, and the definition of PPA. Section~\ref{sec:dna} provides the necessary concepts in genome assembly for readers without bioinformatics background. Section~\ref{sec:our} presents the implementation of our various operations in PPA-assembler. Finally, we report the experimental results in Section~\ref{sec:results}, and conclude this paper in Section~\ref{sec:conclude}.

\section{Pregel Review}\label{sec:pregel}
For ease of presentation, we first define our graph notations. Given a graph $G=(V, E)$, we denote the number of vertices $|V|$ by $n$, and the number of edges $|E|$ by $m$. We also denote the \emph{diameter} of $G$ by $\delta$. If $G$ is undirected, we denote $v$'s neighbors by $\Gamma(v)$ and $v$'s degree by $d(v)=|\Gamma(v)|$. If $G$ is directed, we denote $v$'s in-neighbors (resp.\ out-neighbors) by $\Gamma_{in}(v)$ (resp.\ $\Gamma_{out}(v)$) and $v$'s in-degree (resp.\ out-degree) by $d_{in}(v)=|\Gamma_{in}(v)|$ (resp.\ $d_{out}(v)=|\Gamma_{out}(v)|$). We denote the ID of $v$ by $id(v)$, and use $v$ and $id(v)$ interchangeably.

\vspace{1mm}

\noindent{\bf Computation Model.} Pregel~\cite{pregel} distributes vertices to different machines in a cluster, where each vertex $v$ is associated with its adjacency list (e.g., $\Gamma(v)$) and its attribute $a(v)$. A program in Pregel implements a user-defined {\em compute}(.) function and proceeds in iterations (called {\em supersteps}). In each superstep, each active vertex $v$ calls {\em compute}({\em msgs}), where {\em msgs} is the set of incoming messages sent from other vertices in the previous superstep. In $v$.{\em compute}({\em msgs}), $v$ may process {\em msgs} and update $a(v)$, send new messages to other vertices, and vote to halt (i.e., deactivate itself). A halted vertex is reactivated if it receives a message in a subsequent superstep. The program terminates when all vertices are inactive and there is no pending message for the next superstep. Finally, the results (e.g., $a(v)$) are dumped to HDFS.

Pregel numbers the supersteps, so that a user may access the current superstep number in {\em compute}(.) to decide the proper behavior. Pregel also supports aggregator, a mechanism for global communication. Each vertex can provide a value to an aggregator in {\em compute}(.) in a superstep. The system aggregates those values and makes the aggregated result available to all vertices in the next superstep.

\vspace{1mm}

\noindent{\bf  Our Extensions to Pregel API.} We find the following two API extensions useful in implementing PPA-assembler. Firstly, for two consecutive jobs $j$ and $j'$, we allow $j'$ to directly obtain input from the output of $j$ in memory. In contrast, existing Pregel-like systems require $j$ to first dump its output to HDFS, which is then loaded again by $j'$. Let the vertex class of job $j$ (resp.\ $j'$) be $V_j$ (resp.\ $V_{j'}$), then to enable the direct memory input, users need to define a user-defined function (UDF) {\em convert}($v$) which indicates how to transform an object $v$ of class $V_j$ (processed by $j$) into (zero or more) input objects of class $V_{j'}$ (for job $j'$). After job $j$ finishes, each machine generates a set of objects of type $V_{j'}$ by calling {\em convert}(.) on its assigned vertices of type $V_j$ (which are then garbage collected). Since Pregel+ distributes vertices to machines by hashing vertex ID, the generated objects of type $V_{j'}$ are then shuffled according to their vertex ID, before running job $j'$.

Secondly, the input data may not be in the format of one line per vertex. For example, each line may correspond to one edge, and hence the adjacency list of a vertex can be obtained from multiple lines. To create vertices from such input data, we support a mini-MapReduce procedure during graph loading. Specifically, each line may generate (zero or more) key-value pairs (using UDF {\em map}(.)) where the key is vertex ID, and these key-value pairs are then shuffled according to vertex ID. After each machine receives its assigned key-value pairs, these pairs are sorted by key, so that all pairs with the same key form a group. Finally, each group with key $id(v)$ are processed (using UDF {\em reduce}(.)) to create the input vertex object $v$.

\vspace{1mm}

\noindent{\bf Practical Pregel Algorithm (PPA).} Our prior work~\cite{ppa} defined a class of scalable Pregel algorithms called PPAs, and it designed PPAs for many fundamental graph problems. These PPAs can be used as building blocks to design PPAs for other sophisticated graph problems, such as DBG-based sequencing studied in this paper.
Formally, a Pregel algorithm is called a \textit{balanced practical Pregel algorithm} (BPPA) if it satisfies the following constraints:
\begin{enumerate}
\setlength{\parsep}{0pt}
\setlength{\itemsep}{0pt}
\item {\em Linear space usage:} each vertex $v$ uses $O(d(v))$ (or $O(d_{in}(v)+d_{out}(v))$) space of storage.
\item {\em Linear computation cost:} the time complexity of the $compute(.)$ function for each vertex $v$ is $O(d(v))$ (or $O(d_{in}(v)+d_{out}(v))$).
\item {\em Linear communication cost:} at each superstep, the size of the messages sent/received by each vertex $v$ is $O(d(v))$ (or $O(d_{in}(v)+d_{out}(v))$).
\item {\em At most logarithmic number of rounds:} the algorithm terminates after $O(\log n)$ supersteps.
\end{enumerate}

Constraints~1-3 offers good load balancing and linear cost at each superstep, while Constraint~4 controls the total running time. Note that Constraint~4 includes those algorithms that run for a constant number of supersteps. For some problems, the per-vertex requirements of BPPA can be too strict, and we can only achieve \emph{overall linear space usage, computation and communication cost} (still in $O(\log n)$ rounds). We call a Pregel algorithm that satisfies these constraints simply as a \textit{practical Pregel algorithm} (PPA). The workload skewness problem can be solved using the request-respond API of Pregel+~\cite{da_www}.

We now review two PPAs proposed in~\cite{ppa}, both will be used by PPA-assembler for finding contigs in Section~\ref{ssec:algo}.

\begin{figure}[t]
	\centering
	\includegraphics[width=0.5\columnwidth]{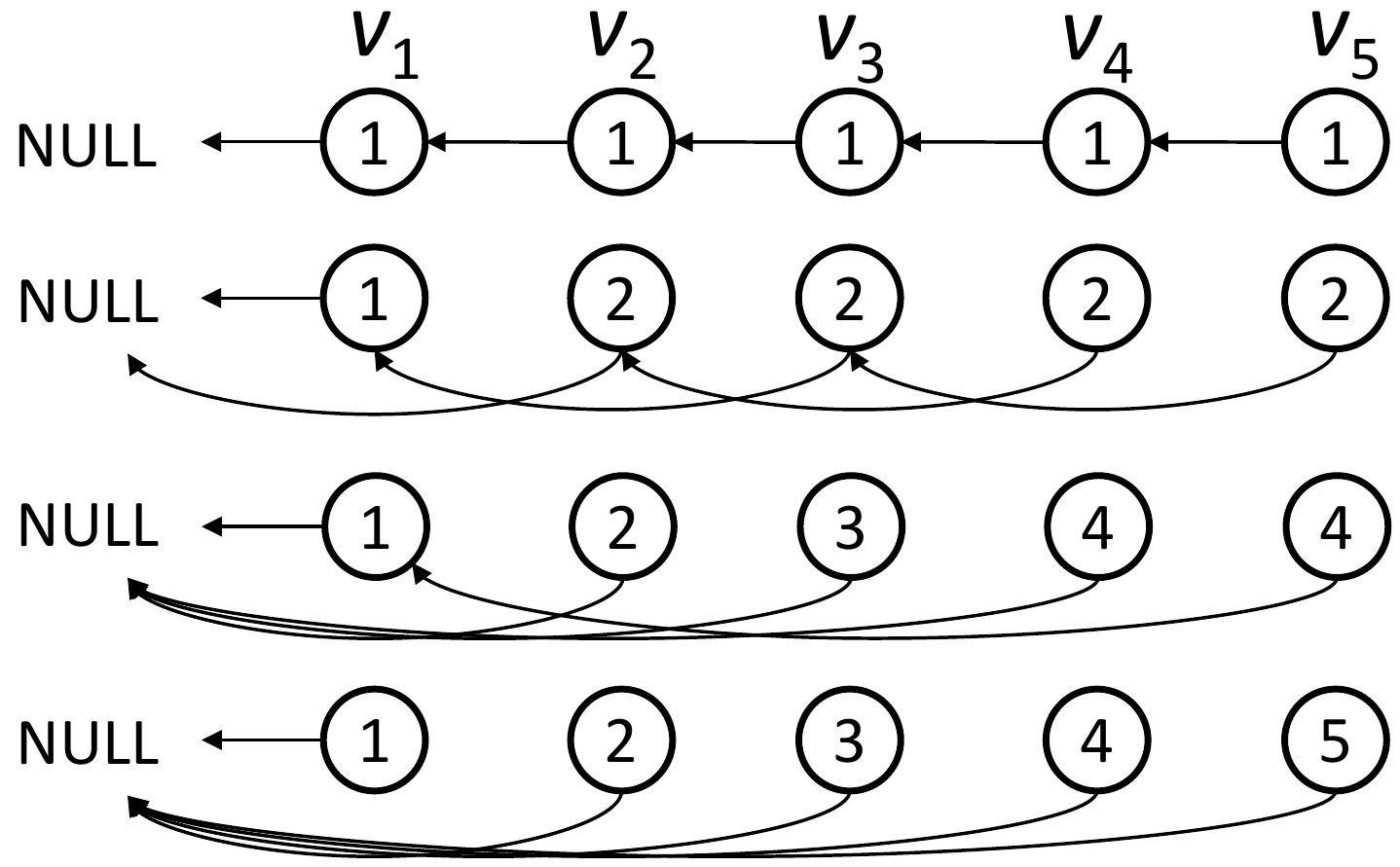}
	\caption{Illustration of BPPA for List Ranking}\label{listRank}
	\vspace{-2mm}
\end{figure}

\vspace{1mm}

\noindent{\em \underline{BPPA for List Ranking.}} Consider a linked list $\mathcal{L}$ with $n$ vertices, where each vertex $v$ keeps a value $val(v)$ and its predecessor $pred(v)$. The vertex $v$ at the head of $\mathcal{L}$ has $pred(v)={\it null}$. For each vertex $v$ in $\mathcal{L}$, let us define $sum(v)$ to be the sum of the values of all the vertices from $v$ following the predecessor link to the head. The \emph{list ranking} problem computes $sum(v)$ for every vertex $v$ in $\mathcal{L}$, where the vertices are stored on HDFS in arbitrary order.

The BPPA for list ranking works as follows. Each vertex $v$ initializes $sum(v)\gets val(v)$. Then in each round, each vertex $v$ does the following in {\em compute}(.): if $pred(v)\neq {\it null}$, $v$ sets $sum(v)\gets sum(v)+sum(pred(v))$ and $pred(v)\gets pred(pred(v))$; otherwise, $v$ votes to halt. Note that to perform these updates, $v$ needs to first request its predecessor $w=pred(v)$ for $sum(w)$ and $pred(w)$, which takes another superstep. This process repeats until $pred(v)={\it null}$ for every vertex $v$, at which point all vertices vote to halt and we have $sum(v)$ as desired.

Figure~\ref{listRank} illustrates how the algorithm works. Initially, objects $v_1$--$v_5$ form a linked list with $sum(v_i)=val(v_i)=1$ and $pred(v_i)$ $=v_{i-1}$. Let us now focus on $v_5$. In Round~1, we have $pred(v_5)=v_4$ and so we set $sum(v_5)\gets sum(v_5)+sum(v_4)=1+1=2$ and $pred(v_5)\gets pred(v_4)=v_3$. One can verify the states of the other vertices similarly. In Round~2, we have $pred(v_5)=v_3$ and so we set $sum(v_5)\gets sum(v_5)+sum(v_3)=2+2=4$ and $pred(v_5)\gets pred(v_3)=v_1$. In Round~3, we have $pred(v_5)=v_1$ and so we set $sum(v_5)\gets sum(v_5)+sum(v_1)=4+1=5$ and $pred(v_5)\gets pred(v_1)={\it null}$. The number of vertices whose values get summed is doubled after each iteration, and thus the algorithm terminates in $\log n$ rounds.

\begin{figure}[t]
	\centering
	\includegraphics[width=0.68\columnwidth]{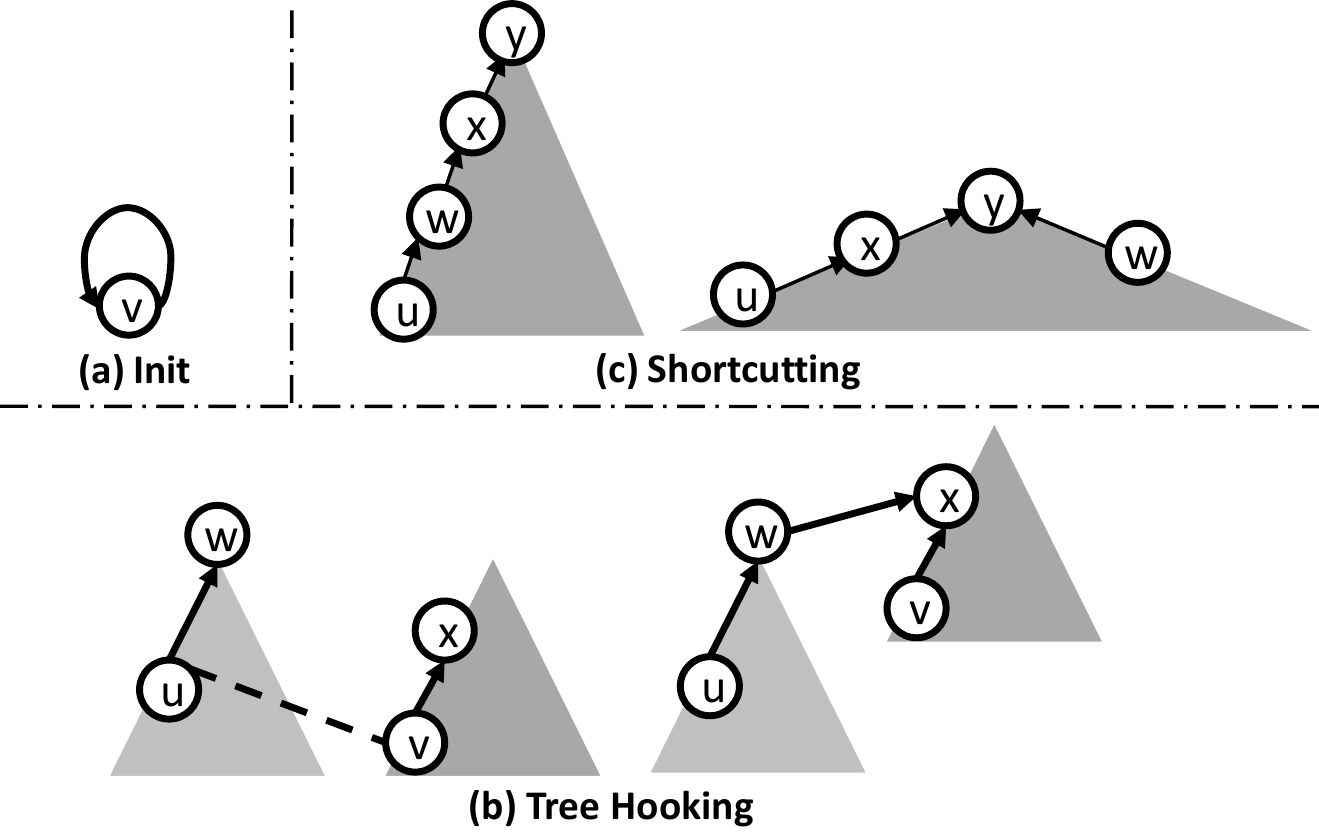}
	\caption{Illustration of the Simplified S-V Algorithm}\label{sv}
	\vspace{-6mm}
\end{figure}

\vspace{1mm}

\noindent{\em \underline{Simplified S-V Algorithm.}} The S-V algorithm was proposed in~\cite{ppa} for computing the connected components (CCs) of a big undirected graph $G$ in $O(\log n)$ number of supersteps, by adapting Shiloach-Vishkin's PRAM algorithm~\cite{shiloach1982logn} to run in Pregel. In the S-V algorithm, each round of computation requires three operations: tree hooking, star hooking, and shortcutting. However, we find that star hooking is actually an artifact required by the original Shiloach-Vishkin's algorithm for correct termination in the PRAM setting. Here, we propose a simplified version of the S-V algorithm that does not require star hooking, which is more efficient as the expensive checking of whether a vertex is in a star (i.e., a tree with height 1) required by the original S-V algorithm is eliminated.

Throughout this algorithm, vertices are organized by a forest such that all vertices in a tree belong to the same CC. Each vertex $v$ maintains a link $D[v]$ to its parent in the forest. We relax the tree definition a bit here to allow the tree root $w$ to have a self-loop (i.e., $D[w]=w$).

At the beginning, each vertex $v$ initializes $D[v]\gets v$, forming a self loop as shown Figure~\ref{sv}(a). Then, the algorithm proceeds in rounds, and in each round, the parent links are updated in two steps: (1)~{\em tree hooking} (see Figure~\ref{sv}(b)): for each edge $(u, v)$, if $u$'s parent $w=D[u]$ is a tree root, we hook $w$ as a child of $v$'s parent $x=D[v]$ (i.e., we merge the tree rooted at $w$ into $v$'s tree); (2)~{\em shortcutting} (see Figure~\ref{sv}(c)): for each vertex $v$, we move it closer to the tree root by linking $v$ to the parent of $v$'s parent, i.e., $D[D[v]]$. Note that Step~2 has no impact on $D[v]$ if $v$ is a root or a child of a root.

The algorithm repeats these two steps until no vertex $v$ has $D[v]$ updated in a round (checked by using aggregator), by which time every vertex is in a star, and each star corresponds to a CC. Since $D[v]$ monotonically decreases during the computation, at the end $D[v]$ equals the smallest vertex in $v$'s CC (which is also the root of $v$'s star). In other words, all vertices with the same value of $D[v]$ constitute a CC. Since each round can be formulated in Pregel as a constant number of supersteps, and shortcutting provides the $O(\log n)$-round bound, the algorithm is a PPA.

\begin{figure}[t]
	\centering
	\includegraphics[width=0.68\columnwidth]{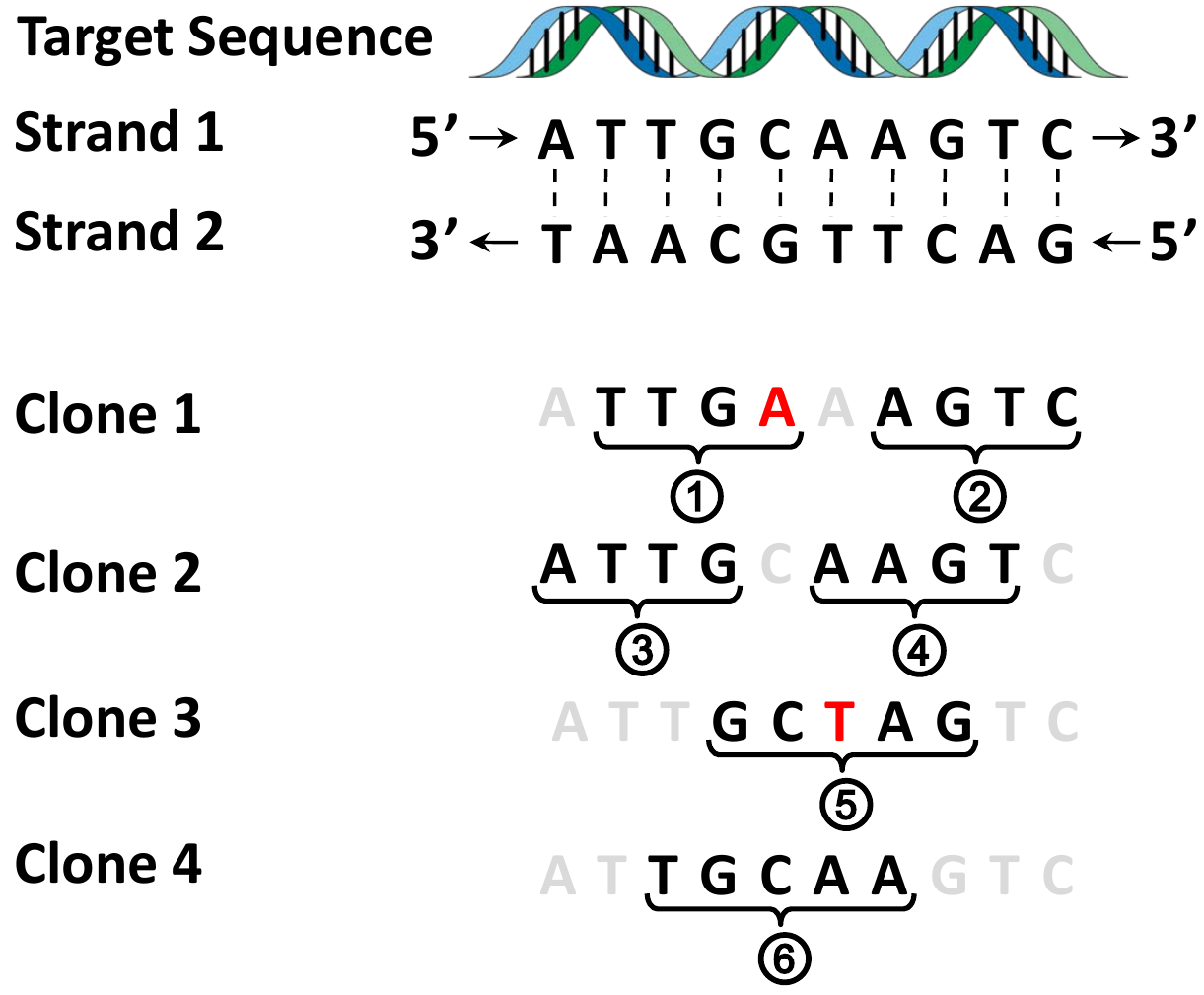}
	\caption{DNA Clones, Reads and Sequencing Errors}\label{clone}
	\vspace{-6mm}
\end{figure}

\section{De Novo Genome Assembly}\label{sec:dna}
This section provides the necessary concepts in DBG-based sequencing for readers without bioinformatics background.

We model a DNA molecule as a very long sequence of nucleotides, where each nucleotide can take one of the four base types A, C, G and T. Since sequencing long DNA segments is error-prone, modern sequencing technologies generate a large number of short DNA segments, called {\em reads}. Figure~\ref{clone} illustrates this process, where 4 DNA clones are sheared into 6 reads. Note that a DNA molecule consists of two strands coiled around each other, and we only consider strand~1 in Figure~\ref{clone} for simplicity. Reads can have variable lengths, and sequencing errors may happen at some positions such as in reads~\textcircled{1} and~\textcircled{5} in Figure~\ref{clone} (errors highlighted in red). Also, reads may overlap with each other, such as reads~\textcircled{2} and~\textcircled{4} in Figure~\ref{clone} that share the segment ``AGT''. It is through these overlaps that genome assembly algorithms stitch reads to get longer sequences (called {\em contigs}) or even the whole sequence.

\begin{figure}[t]
	\centering
	\includegraphics[width=\columnwidth]{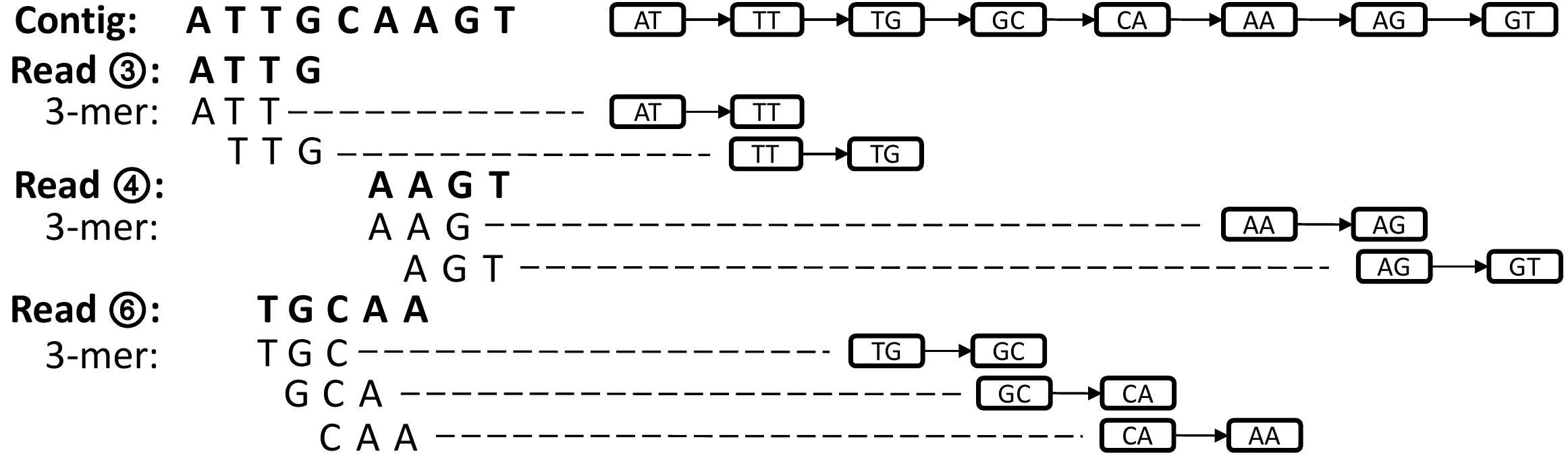}
	\caption{Illustration of $k$-mers}\label{kmer}
	\vspace{-6mm}
\end{figure}

\vspace{1mm}

\noindent{\bf De Bruijn Graph \& $k$-mer.} The DBG-based assembly approach first constructs a de Bruijn graph (DBG) from the reads, and then find contigs from the DBG. To construct a DBG, each read is cut into consecutive sub-sequences of length $k+1$, where each sub-sequence is called a $(k+1)$-mer. For example, Figure~\ref{kmer} illustrates how we can generate 3-mers from reads~\textcircled{3}, \textcircled{4} and~\textcircled{6} of Figure~\ref{clone} (we consider $k=2$ here), where read~\textcircled{3} ``ATTG'' can be cut into two 3-mers ``ATT'' and ``TTG''. For each $(k+1)$-mer, we define its {\em prefix} (resp. {\em suffix}) as the subsequence without the last (resp.\ first) nucleotide, which is a $k$-mer. The $k$-mers define the vertices in the DBG, and each $(k+1)$-mer defines an edge from its prefix to its suffix in the DBG. For example, in Figure~\ref{kmer}, the first 3-mer of read~\textcircled{3}, i.e., ``ATT'', defines a directed edge in DBG from vertex~``AT'' to vertex~``TT''. From all the 3-mers in Figure~\ref{kmer}, we can create a path as shown at the top right corner of Figure~\ref{kmer}, which essentially stitches reads~\textcircled{3}, \textcircled{4} and~\textcircled{6} together into a longer contig ``ATTGCAAGT''.

Ideally, we would like to choose $k$ to be large enough so that any sub-sequence of length $k$ in the whole DNA sequence appears only once, i.e., any $k$-mer vertex of the DBG corresponds to a unique sub-sequence in the whole sequence. If this is the case (an extreme case is when we set $k$ equal to the length of the whole sequence), the DBG is essentially a path like in Figure~\ref{kmer}, following which we can reconstruct the whole sequence. Also, since we are essentially hashing the $k$-mers contained in the whole sequence to $4^k$ possible values (4 is because each nucleotide can take four values), a larger $k$ reduces the chance of having two DNA sub-sequences of length $k$ with the same sequence content. However, we cannot set $k$ arbitrarily large, since reads are short and any read with length less than $(k+1)$ cannot contain any $(k+1)$-mer and will be ignored. We should set $k$ to use the majority of the reads when constructing the DBG, but a $k$-mer vertex in the DBG may correspond to several different segment positions in the whole sequence. We call such a vertex as {\em ambiguous}. It is clear that the whole sequence corresponds to a Eulerian path of the DBG (not considering long repetitive sequence patterns that cause cycles), but there can be many Eulerian paths in the DBG. Thus, the goal of DBG-based sequencing is to find the maximal simple paths in the DBG that do not contain any ambiguous vertex, which constitute contigs which are longer sub-sequences of the whole sequence.

\vspace{1mm}

\begin{figure}[t]
	\centering
	\includegraphics[width=0.9\columnwidth]{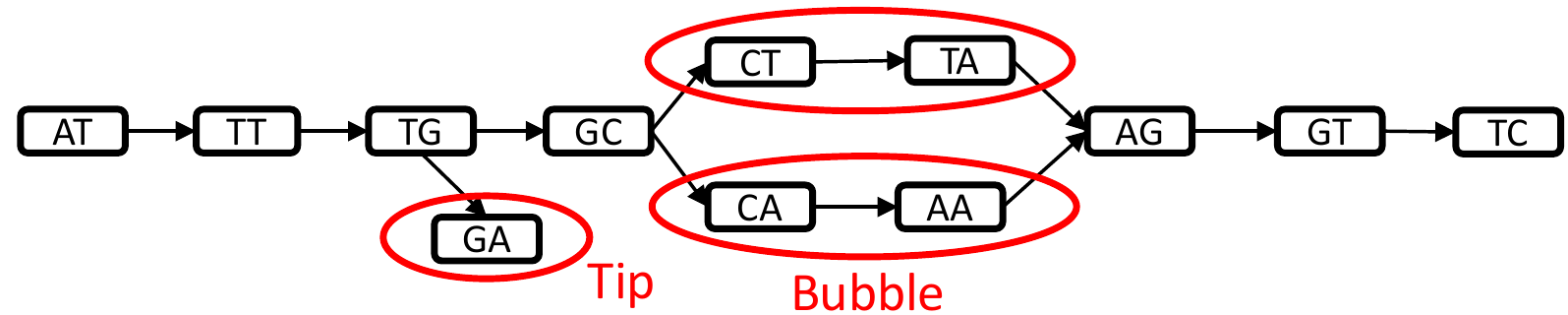}
	\caption{De Bruijn Graph}\label{dbg}
	\vspace{-6mm}
\end{figure}

\noindent{\bf Error Correction.} The above sequencing method does not consider the possibility that reads may contain errors. Read errors can further complicate the assembly process by introducing false vertices and edges into the DBG. Two typical errors are {\em tips} and {\em bubbles}, as illustrated in Figure~\ref{dbg} which shows the DBG constructed from the reads of Figure~\ref{clone}. A {\em tip} is a short dangling path in the DBG that leads to a dead-end, such as edge ``TG''$\rightarrow$``GA'' in Figure~\ref{dbg} that is contributed by the error in read~\textcircled{1}. A {\em bubble} is a sub-path that starts from a certain vertex at the main path of the DBG, and returns to the same path after a few hops. Figure~\ref{dbg} shows a bubble where the main path ``GC''$\rightarrow$``CA''$\rightarrow$``AA''$\rightarrow$``AG'' is contributed by correct reads such as \textcircled{4} and~\textcircled{6}, and the erroneous sub-path ``GC''$\rightarrow$``CT''$\rightarrow$``TA''$\rightarrow$``AG'' is caused by read~\textcircled{5} that has an error. Note that vertex ``TG'', ``GC'' and ``AG'' become ambiguous simply because of the read errors. If we can correct the errors, we can obtain longer contigs. For example, in Figure~\ref{dbg}, we essentially reconstruct the whole sequence as a contig after removing erroneous paths.

However, we should not be overly aggressive when correcting potential errors, as false alarms may create wrong (albeit longer) contigs. For example, we only consider a dangling path as a tip if it is short, as a long tip needs to be generated by multiple errors which is unlikely. In fact, a long dangling path is mostly likely to be a valid contig, with its dead-end caused by no read coverage at the corresponding position in the whole sequence (or covered by reads with length less than $(k+1)$). Here, we define the {\em coverage of a base-pair} (or, position) in the whole sequence as the number of reads that covers it. For example, in Figure~\ref{clone}, the first nucleotide has coverage 1 (covered only by read~\textcircled{3}), and the second nucleotide has coverage 2 (covered by reads~\textcircled{1} and~\textcircled{3}). As there are many DNA clones, it is unlikely (but still possible) that a particular base-pair is never covered.

As for a bubble, we remove sub-path(s) with a very low coverage. Here, we define the {\em coverage of a path} as the minimum coverage among all edges on the path, where the {\em coverage of an edge} (i.e., a $(k+1)$-mer) is defined as the number of reads that generate that $(k+1)$-mer. A correct path is unlikely to have a low coverage as there are many DNA clones, and a low coverage is often contributed by an erroneous read. We also require a sub-path to be similar to the main path (with high coverage) in order to remove it, since it is unlikely to have multiple errors that significantly changes the corresponding sub-sequence. The similarity can be measured by the edit distance between the two sub-sequences represented by these two paths.

\vspace{1mm}

\noindent{\bf Directionality.} In the discussion so far, we have been assuming that the DNA molecule is just a long sequence. In reality, reads may be obtained from both strands of the DNA molecule. As Figure~\ref{clone} on Page~3 shows, each strand has an end-to-end chemical orientation, and reads are always obtained in the 5'-to-3' direction. Specifically, strand~1 (resp.\ strand~2) is read from left to right (resp.\ from right to left). Let us temporarily read both strands from left to right, then two nucleotides at the same position in both strands constitute a {\em base-pair}. For example, in Figure~\ref{clone}, the first two base-pairs are $(A, T)$ and $(T, A)$. Nucleotides $A$ and $T$ (resp.\ $G$ and $C$) are complementary to each other, and the two nucleotides in a base-pair are always complementary, as shown in Figure~\ref{clone}.

Given a nucleotide $x$, we denote its complement by $\overline{x}$. The {\em reverse complement} of a DNA sequence $s=x_1x_2\ldots x_\ell$ is denoted by $rc(s)=\overline{x_\ell}\ \overline{x_{\ell-1}}\ldots\overline{x_1}$ (or simply $\overline{x_\ell x_{\ell-1}\ldots x_1}$). For example, the reverse complement of strand~1 in Figure~\ref{clone} is ``GACTTGCAAT'', which is exactly strand~2 reading in the 5'-to-3' direction.

\begin{figure}[t]
	\centering
	\includegraphics[width=0.8\columnwidth]{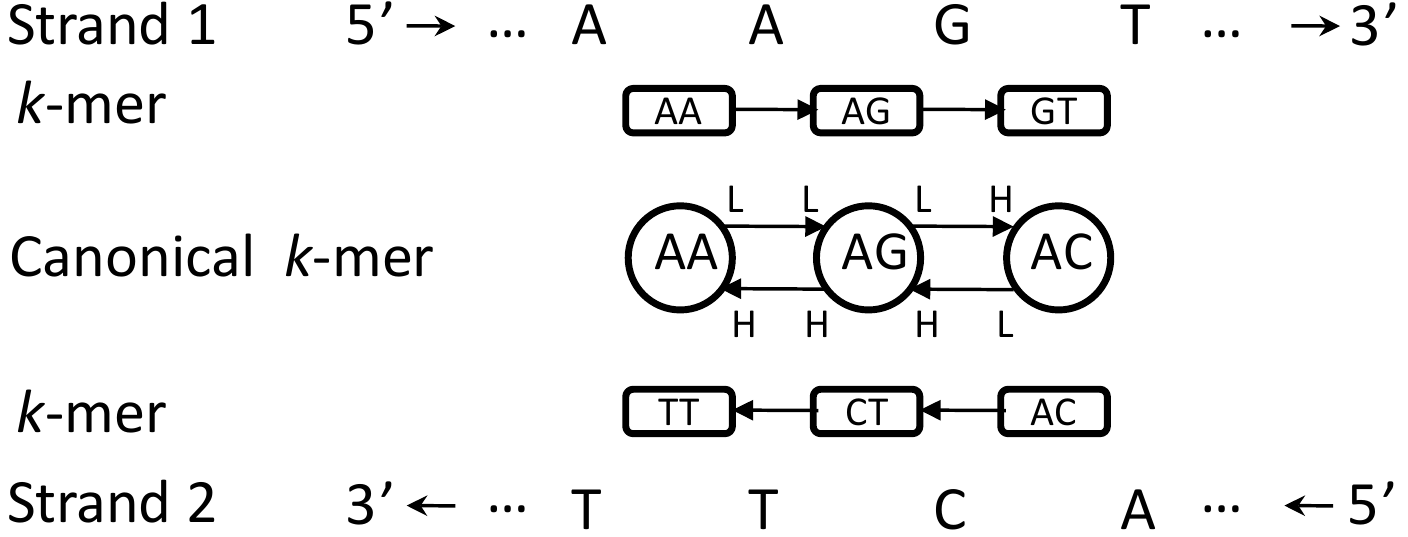}
	\caption{Canonical $k$-mers \& Edge Polarity}\label{rc}
	\vspace{-6mm}
\end{figure}

Now consider read~\textcircled{4} in Figure~\ref{clone} which is re-plotted on strand~1 in Figure~\ref{rc}. If we read the same DNA segment on strand~2 in the 5'-to-3' direction, we obtain another read ``ACTT'', which is exactly the reverse complement of read~\textcircled{4}. Figure~\ref{rc} also shows the $k$-mer vertices and $(k+1)$-mer edges generated by these two reads ($k=2$).

Ideally, we would like any length-$k$ sub-sequence of a strand to appear only once in that strand, and not to appear in the other strand, so that each $k$-mer decides a unique segment location in only one of the two strands. Assume this is the case, it is not difficult to see in Figure~\ref{rc} that a $k$-mer and its reverse complement refer to the same position in the DNA molecule. For example, the two rightmost $k$-mer nodes ``GT'' and ``AC'' are the reverse complement of each other. We would like a $k$-mer and its reverse complement to correspond to a unique vertex in the DBG, so that reads from different strands can be stitched to create longer contigs as long as the reads share overlapping DNA segments. To achieve this goal, we define the {\em canonical $k$-mer} of a $k$-mer $s$ as the lexicographically smaller sequence between $s$ and $rc(s)$, and use the canonical $k$-mer as a vertex in the DBG. For example, the rightmost $k$-mers ``GT'' and ``AC'' in Figure~\ref{rc} both refer to the rightmost DBG vertex ``AC'' of the chain in the middle of Figure~\ref{rc}.

Accordingly, now each DBG edge $(u, v)$ needs to have a {\em polarity} to indicate the direction of a $(k+1)$-mer that generates this edge, i.e., $u$-to-$v$ or $v$-to-$u$. Polarity is used to indicate the stitching directions when constructing contigs. We use an example to explain how edge polarity is determined. Consider the last $(k+1)$-mer of read ``AAGT'' from strand~1 in Figure~\ref{rc} ($k=2$), i.e., ``AGT'', which creates an edge ``AG''$\rightarrow$``GT''. Edge source ``AG'' is already canonical and thus we give it a label $L$, while edge target ``GT'' needs to be converted to its reverse complement ``AC'' to be a DBG vertex, in which case we give it a label $H$. The edge direction is simply a concatenation of the source and target labels, i.e., $\langle L:H\rangle$.

We say that labels $H$ and $L$ are complementary, and denote $\overline{H}=L$ and $\overline{L}=H$. It is not difficult to see the following property (e.g., from Figure~\ref{rc}).

\begin{property}\label{prop}
Edge $(u, v)$ with polarity $\langle X:Y\rangle$ is equivalent to edge $(v, u)$ with polarity $\langle \overline{Y}:\overline{X}\rangle$.
\end{property}

This property allows us to stitch $k$-mers generated from different strands. For example, consider $(k+1)$-mers ``AAG'' from strand~1 and ``ACT'' from strand~2, which generates two edges ``AA''$\xrightarrow{\langle L:L\rangle}$``AG'' and ``AC''$\xrightarrow{\langle L:H\rangle}$``AG''. Although both edges are incident to ``AG'', the labels at the side of ``AG'' do not match. Since the latter edge is equivalent to ``AG''$\xrightarrow{\langle L:H\rangle}$``AC'', we can stitch the edges to obtain ``AA''$\xrightarrow{\langle L:L\rangle}$``AG''$\xrightarrow{\langle L:H\rangle}$``AC'' where both edges agree on label $L$ for ``AG'' and are in the same direction.

Finally, we remark that our discussion has been assuming the ideal case, and in reality a $k$-mer may appear in multiple positions in both strands. Such a $k$-mer is ambiguous, and our goal is still to find the contigs, i.e., the maximal simple paths in the DBG that do not contain any ambiguous vertex.

\section{PPA-Assembler Algorithms}\label{sec:our}
We first present our compact graph data structures, and then describe the operations supported by PPA-Assembler.

\subsection{Vertex \& Edge Formats}\label{ssec:ds}
We design compact data structures for vertices and edges in our vertex-centric programs to be {\bf memory-efficient} (note that genome assembly has a very high memory demand~\cite{kleftogiannis2013comparing}).

\begin{figure}[t]
	\centering
	\includegraphics[width=0.76\columnwidth]{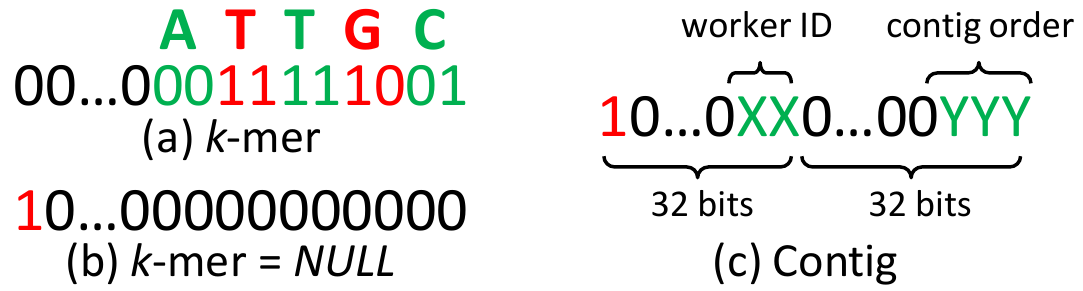}
	\caption{Vertex ID Format}\label{id}
	\vspace{-6mm}
\end{figure}

\vspace{1mm}

\noindent{\bf Vertex ID.} Each vertex in a Pregel program has a unique ID for message passing, and we use integer to specify vertex ID. There are two kinds of vertices in PPA-Assembler, (1)~{\bf $k$-mer} and (2)~{\bf contig}. We encode the sequence of a $k$-mer directly into its integer ID, so that different $k$-mers have different IDs. Recall that reads are cut into $(k+1)$-mers during DBG construction. Without loss of generality, let us assume that $k\leq 31$, and hence we use 64-bit integer for ID (more bits will be used if $k>31$). Each nucleotide is represented by two bits: A (00), T (11), G (10), C (01), and thus a $k$-mer requires at most 62 bits to represent. We align this binary sequence to the right of the 64-bit ID, and pad all remaining bits (at least 2) on the left with zeros. As an example, Figure~\ref{id}(a) shows the ID of a 5-mer ``ATTGC''. Sometimes, we would like to indicate that a $k$-mer or a contig has no neighbor along one direction (e.g., the dead-end of a tip). We use a dummy neighbor in this case, denoted by {\em NULL}, whose ID is given by the special 64-bit binary sequence with the most significant bit being 1 and all others being 0. Finally, since a contig can be an arbitrarily long sequence, we cannot encode the sequence into the contig's ID. Instead, since the contigs are distributed among the machines after their generation in PPA-assembler, we let the $i$-th worker machine assign its $j$-th contig a 64-bit ID that equals the 32-bit integer representation of $i$ concatenated with the 32-bit integer representation of $j$ as shown in Figure~\ref{id}(c). To avoid collision with the ID of a $k$-mer, we also flip the most significant bit to be 1.

Compared with directly using sequence as vertex ID, using integer ID provides the following benefits: (1)~Pregel heavily checks vertex IDs for message delivery, and integer IDs benefit from efficient word-level instructions; (2)~no additional space is needed to store the sequence of a $k$-mer vertex, and the sequence-related processing can be efficiently executed by bitwise operations.

\vspace{1mm}

\begin{figure}[t]
	\centering
	\includegraphics[width=\columnwidth]{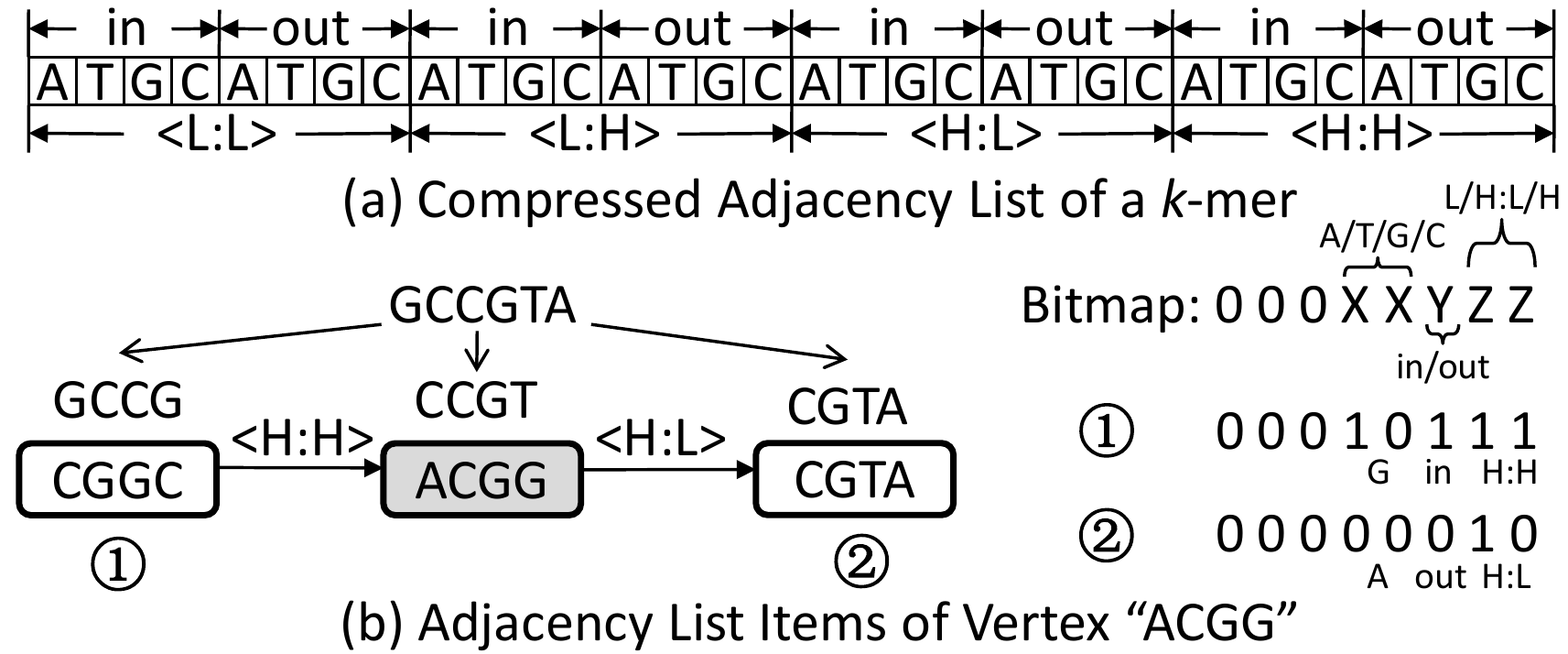}
	\caption{Adjacency List Format}\label{bitmap}
	\vspace{-6mm}
\end{figure}

\noindent{\bf Format of a $k$-mer Vertex.} Each vertex also maintains an adjacency list of its neighboring vertices ($k$-mers or contigs). Since a contig is obtained by merging unambiguous $k$-mers, it has only two neighbors along its two opposite sequencing directions, where each neighbor is either an ambiguous $k$-mer or {\em NULL} (i.e. the dead-end).

In contrast, a $k$-mer vertex may have more than two neighbors. There are two cases: (1)~at the beginning, all vertices (and hence all neighbors) are $k$-mers that are generated from reads; (2)~in later processing, a neighbor of a $k$-mer may also be a contig that is generated by merging unambiguous $k$-mers (this later processing makes sense since error correction may render the current $k$-mer unambiguous, leading to further contig merging). Case~(1) is the most memory-consuming since the overlapping $k$-mers incur a lot of data redundancy; once unambiguous $k$-mers (which are often the majority) get merged into contigs, the memory consumption is usually no longer a problem due to the significantly reduced data volume.

Therefore, we compress the adjacency lists of $k$-mer vertices in Case~(1) using compact bitmaps, which we describe next. Let us first ignore edge polarity, then a $k$-mer can have at most 4 in-neighbors and 4 out-neighbors (i.e., 8 neighbors). For example, the 4-mer ``CCGT'' can have at most 4 in-neighbors whose suffix matches its prefix ``CCG'', i.e., ``ACCG'', ``TCCG'', ``GCCG'' and ``CCCG''. Now taking the 4 possible edge polarity $\langle L:L\rangle$, $\langle L:H\rangle$, $\langle H:L\rangle$ and $\langle H:H\rangle$ into account, we obtain $4\times 8=32$ possible combinations, which we represent using the 32-bit bitmap shown in Figure~\ref{bitmap}(a). A bit is 1 if the corresponding neighbor exists (and it is 0 otherwise). For example, if the rightmost bit is 1, then we can obtain the neighbor (i.e., its ID that encodes the sequence) by (i)~reverse-complementing the current $k$-mer (i.e., its ID) since the left half of edge polarity is $H$, (ii)~appending ``C'' to its suffix, and (iii)~reverse-complementing the resulting sequence since the right half of edge polarity is $H$. In addition to the 32-bit neighbor bitmap, a $k$-mer vertex also maintains a list of counts, one for each neighbor (i.e., each bit 1 in the bitmap) which records the coverage of the corresponding edge. The counts are stored as variable-length integers to save space (e.g., a small count can often be represented with just one byte). In fact, using Property~\ref{prop} mentioned in Section~\ref{sec:dna}, we can actually further cut the bitmap size by half.

In uncompressed format, each $k$-mer neighbor (i.e., adjacency list item) of a $k$-mer vertex is represented by an 8-bit bitmap plus the coverage of the corresponding edge. The bitmap format is shown in Figure~\ref{bitmap}(b), where the leftmost three bits are always 0, two bits $XX$ are used to indicate what nucleotide gets prepended (resp.\ appended) to the prefix (resp.\ suffix) of the current $k$-mer vertex to form the neighbor, one bits $Y$ is used to indicate whether the neighbor is an in-neighbor or an out-neighbor, and two bits $ZZ$ are used to indicate the polarity of the corresponding edge. For example, consider the 4-mer vertex ``ACGG'' in Figure~\ref{bitmap}(b). Its in-neighbor, node~\textcircled{1}, is represented by the bitmap 00010111 which indicates that the edge polarity is $\langle H:H\rangle$, and that the neighbor's sequence ``CGGC'' is obtained by reverse-complementing ``ACGG'' into ``CCGT'', prepending G (10) to the prefix ``CCG'' to obtain ``GCCG'', and then reverse-complementing it into ``CGGC''. Finally, sometimes we need to indicate that a $k$-mer vertex reaches a dead-end along one direction, in which case we use the bitmap 10000000 to indicate that the neighbor is {\em NULL}.

A $k$-mer vertex tracks its contig neighbors differently from the $k$-mer neighbors, and we shall discuss the format shortly.

\vspace{1mm}

\begin{figure*}[t]
	\centering
	\includegraphics[width=1.3\columnwidth]{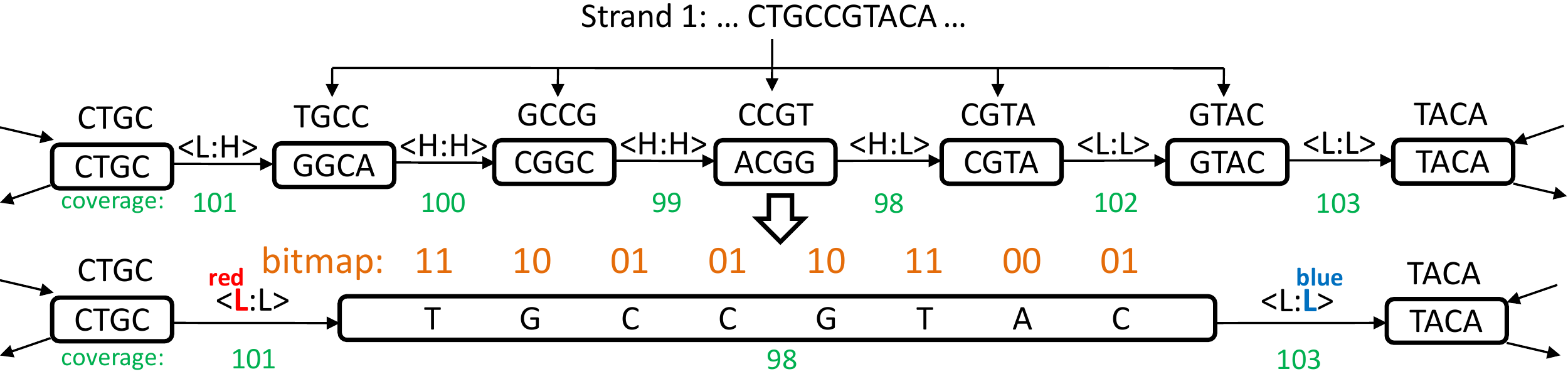}
	\caption{Contig Format}\label{contig}
	\vspace{-6mm}
\end{figure*}

\noindent{\bf Format of a Contig.} Recall from Figure~\ref{id}(c) that the ID of a contig does not contain its sequence information. In PPA-assember, a contig vertex keeps its sequence as a variable-length bitmap. For example, Figure~\ref{contig} shows a DBG path where only the two end $k$-mer vertices are ambiguous, and the other $k$-mers are merged into a contig with bitmap 11\,10\,01\,01\,10\,11\,00\,01. We always keep the contig-side edge polarity to be $L$, so that the contig corresponds to the sequence in strand~1 rather than strand~2. Besides the bitmap, a contig vertex also maintains an in-neighbor (resp.\ out-neighbor) such as the $k$-mer vertex ``CTGC'' (resp.\ ``TACA'') in Figure~\ref{contig}. Note that the in-neighbor and out-neighbor are uniquely defined since we already specify the sequencing direction of any contig, i.e., 5'-to-3' in strand~1. Besides ID, each neighbor is also stored with the neighbor-side edge polarity (e.g., the red and blue $L$'s in Figure~\ref{contig}), and the coverage of the corresponding adjacent edge (e.g., 101 and 103 in Figure~\ref{contig}). Finally, a contig vertex also maintains its own coverage (e.g., 98 in Figure~\ref{contig}), which is computed as the minimum coverage of all edges (i.e., $(k+1)$-mers) merged by the contig.

We now consider how a $k$-mer vertex $v_{me}$ tracks its contig neighbors. Recall that $v_{me}$ tracks each $k$-mer neighbor by a bitmap plus a coverage. It tracks each contig neighbor $v_{contig}$ quite differently. Referring to Figure~\ref{contig} again, while we can view the contig as the neighbor of vertex ``CTGC'' (and ``TACA''), another perspective is to treat it as a label on the edge connecting ``CTGC'' to ``TACA''. We adopt the latter perspective, and let $v_{me}$ maintain $v_{contig}$'s information as a triplet including (i)~the $k$-mer vertex's ID on the other end of $v_{contig}$, denoted by $v_{other}$; (ii)~direction of edge $(v_{me}, v_{other})$ (incoming or outgoing) and polarity (e.g., the red and blue $L$'s in Figure~\ref{contig}); (iii)~$v_{contig}$'s ID which can be used by $v_{me}$ to request for $v_{contig}$'s sequence (e.g., for further contig merging). Other information about $v_{contig}$ such as sequence length and coverage can also be materialized in its corresponding adjacency list item of $v_{me}$ to facilitate tip removing and bubble filtering, which eliminates the cost of requesting them from $v_{contig}$ during processing.

\vspace{1mm}

\noindent{\bf Vertex Types.} First consider a $k$-mer vertex $v$, and it can be of one of the following three types: (1)~$\langle1\rangle$: such a vertex only has one neighbor, and is thus a dead-end; (2)~$\langle1$-$1\rangle$: such a vertex has two neighbors, and when both edges agree on the polarity label for $v$ (either $L$ or $H$) which can be enforced using Property~\ref{prop}, one neighbor is an in-neighbor and the other is an out-neighbor; such a vertex is unambiguous; (3)~$\langle m$-$n\rangle$: such a vertex has at least two neighbors, but it does not satisfy the requirement of $\langle1$-$1\rangle$; such a vertex is ambiguous. Note that $v$ cannot have no neighbor, since a $k$-mer vertex is contributed by the prefix or suffix of a $(k+1)$-mer.

Since a contig is generated by merging unambiguous $k$-mers, it can only be of type $\langle1\rangle$ or type $\langle1$-$1\rangle$. Here, we say that a contig vertex is of type $\langle1\rangle$ iff at least one of its two neighbors is {\em NULL}, i.e., the contig corresponds to a dangling path in DBG and is thus a tip candidate. Note that it is possible to have an isolated contig where both ends are dead-ends (i.e., {\em NULL}), and will be regarded as a tip unless it is long.

\subsection{Operations and Their Algorithms}\label{ssec:algo}
PPA-assembler provides a library of operations for flexible genome assembly in a distributed environment deployed with Hadoop. Each operation is implemented as a PPA (described in Section~\ref{sec:pregel}) and is thus scalable; it can either load data from HDFS, or directly obtain input from another operation's output in memory. Users may combine the provided operations to implement various sequencing strategies, and they may even integrate new operations or redefine existing operations (e.g., changing the criteria for judging tips and bubbles) using Pregel+'s vertex-centric API.

\vspace{2mm}

\begin{figure}[t]
	\centering
	\includegraphics[width=\columnwidth]{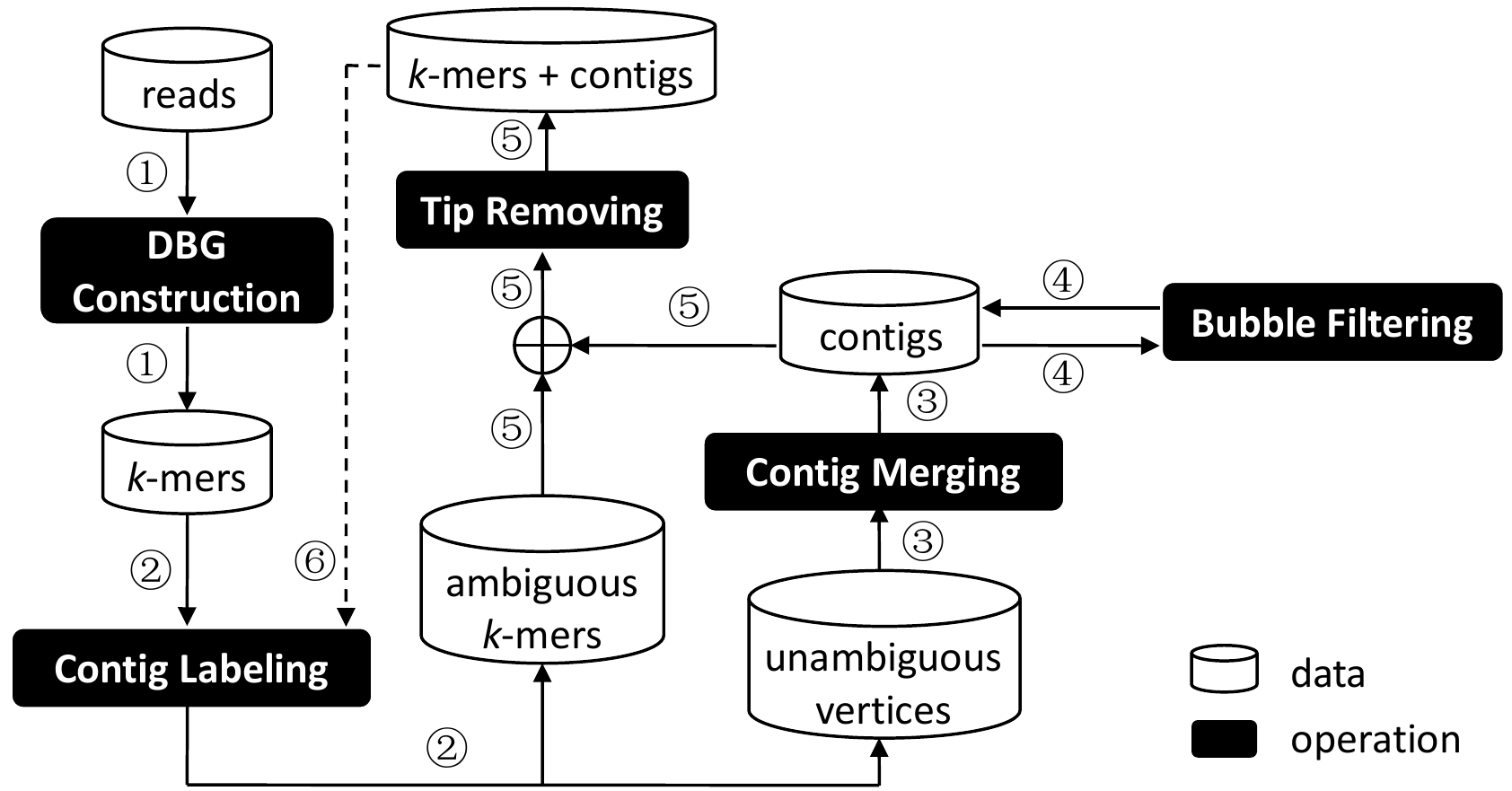}
	\caption{Operation Diagram}\label{overview}
	\vspace{-6mm}
\end{figure}

\begin{figure*}[t]
	\centering
	\includegraphics[width=1.3\columnwidth]{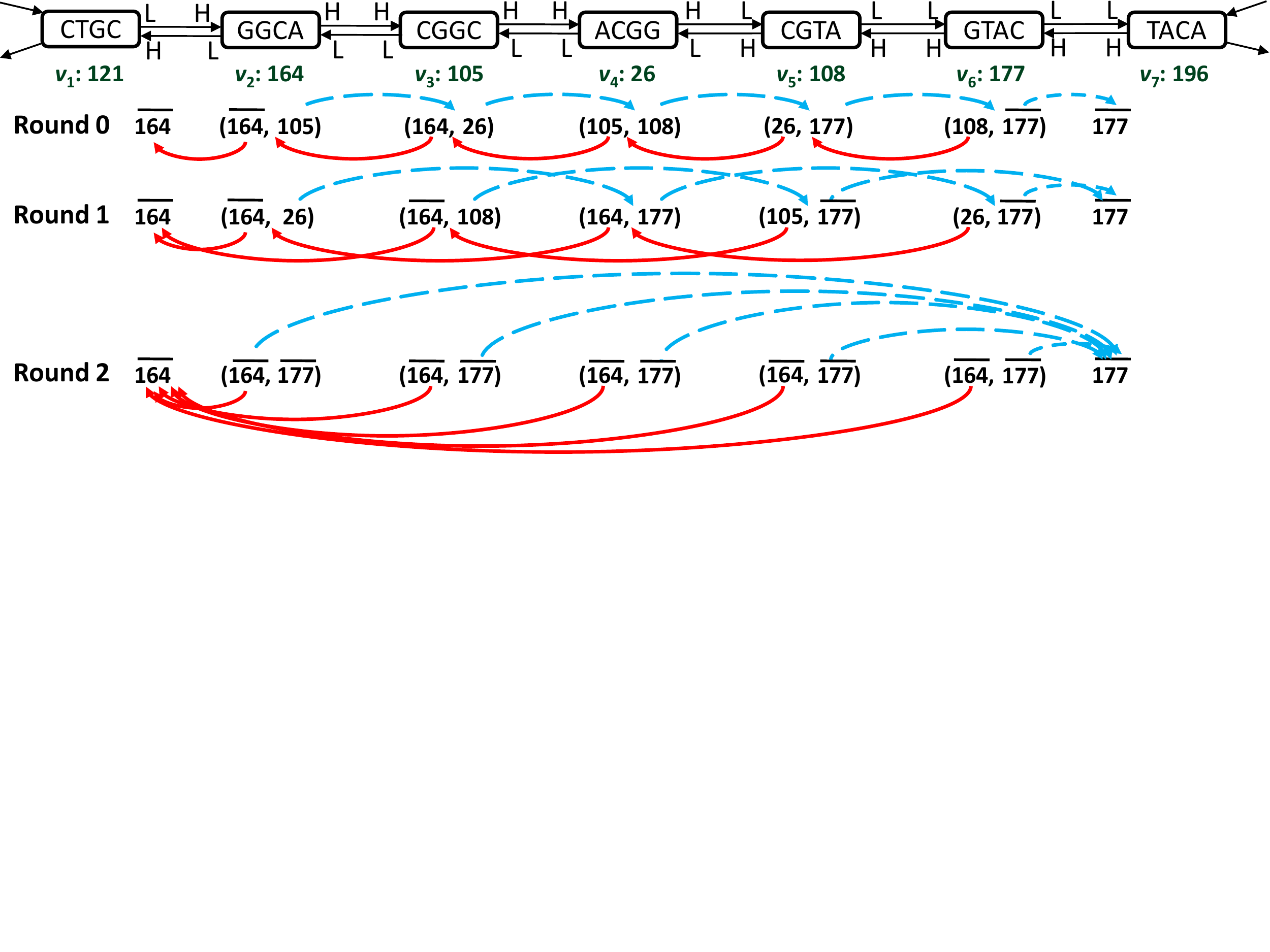}
	\caption{Bidirectional List Ranking}\label{biLR}
	\vspace{-6mm}
\end{figure*}

\noindent{\bf Overview.} Figure~\ref{overview} shows the data flow diagram of PPA-assembler, which includes five operations: \textcircled{1}~{\em DBG construction}, which constructs a DBG from the DNA reads, and outputs the $k$-mer vertices of the DBG along with their adjacency lists; \textcircled{2}~{\em contig labeling}, which divides the vertices into two sets (ambiguous ones and unambiguous ones) and labels unambiguous vertices by the contigs that they belong to; \textcircled{3}~{\em contig merging}, which merges unambiguous vertices into contigs according to the labels; \textcircled{4}~{\em bubble filtering}, which filters any low-coverage contig that shares both ends with another high-coverage contig that have a similar sequence; \textcircled{5}~{\em tip removing}, which takes the ambiguous $k$-mers and the contigs (after bubble filtering), and removes tips.

In fact, the output of tip removing can be fed to the ``contig labeling'' operation again to grow longer contigs (see arrow \textcircled{6} in Figure~\ref{overview}), since the previous error correction operations may have converted some ambiguous $k$-mer vertices into unambiguous ones, and the operations \textcircled{2}--\textcircled{5} may loop as many times as needed (though we typically just loop for one more round). At the first round, the inputs to operations ``contig labeling'' an ``contig merging'' must be $k$-mers, but starting from the second round, the inputs may contain a mix of $k$-mers and contigs. For ease of discussion, we focus on the first round when discussing operations ``contig labeling'' an ``contig merging''.

\vspace{1mm}

\noindent{\bf \textcircled{1} DBG Construction.} This operation loads DNA reads from HDFS, and creates a DBG from them through two mini MapReduce phases: (i)~the first phase extracts $(k+1)$-mers from reads, and (ii)~the second phase constructs $k$-mer vertices and their adjacency lists from the extracted $(k+1)$-mers, which form the DBG.

We first describe phase~(i). In real DNA data, a read's sequence may contain element ``N'' besides ``A'', ``T'', ``G'', ``C'', and such an element indicates that the nucleotide cannot be determined due to noise in measurement. For this purpose, in {\em map(.)}, a read is first split into sequences by elements ``N'', and each sequence is parsed to obtain the $(k+1)$-mers using a sliding window of $(k+1)$ elements as illustrated in Figure~\ref{kmer}. The sequence of a $(k+1)$-mer is directly encoded in its 64-bit integer ID, which functions as the key for shuffling. In each worker machine, if a $(k+1)$-mer is obtained for the first time, the worker creates an $(ID, count)$ pair for it where $count=1$; otherwise, the $(k+1)$-mer's count is increased by 1. After shuffling, for each $(k+1)$-mer, all its counts (from all workers) are input to {\em reduce(.)}, which then sums these counts to obtain the total count of the $(k+1)$-mer; {\em reduce(.)} only outputs the $(k+1)$-mer as an $(ID, count)$ pair, if the coverage $count>\theta$ where $\theta$ is a user-defined threshold. We filter a low-coverage $(k+1)$-mer since it is very likely to be contributed by erroneous readers.

In Phase~(ii), each remaining $(k+1)$-mer is input to {\em map(.)}, which extracts two $k$-mers that correspond to its prefix and suffix. In each worker, if an extracted $k$-mer is obtained for the first time, the worker creates a $k$-mer vertex for it. A directed edge from the prefix $k$-mer vertex to the suffix $k$-mer vertex is also added into the adjacency lists of both $k$-mer vertices, where an adjacency list is represented by a 32-bit bitmap as shown in Figure~\ref{bitmap}, and edge count (which equals the $(k+1)$-mer's count) is also recorded or incremented, using a variable-length integer. The $k$-mer vertices with partially constructed adjacency lists are then shuffled by the 64-bit integer ID. After shuffling, for each $k$-mer, its partial adjacency lists (from all workers) are input to {\em reduce(.)}, which combines them to obtain the complete adjacency list (still represented in 32 bits), and sums the counts of the each edge to obtain the edge coverage (represented compactly using a variable-length integer).

\vspace{1mm}

\noindent{\bf \textcircled{2} Contig Labeling.} Let us call a path that only contains vertices of types $\langle1\rangle$ and $\langle1$-$1\rangle$ as an unambiguous path. The ``contig labeling'' operation marks all vertices on each maximal unambiguous path with a unique label, so that they can be grouped to create a contig later. The operation is executed right after ``\textcircled{1} DBG construction'', and the input vertices are all $k$-mers. It can also be executed after ``\textcircled{5} tip removing'' to find longer contigs, in which case some input vertices could already be contigs.


A vertex is at one end of a maximal unambiguous paths, if its type is $\langle1\rangle$, or if its type is $\langle1$-$1\rangle$ and at least one neighbor is of type $\langle m$-$n\rangle$. For example, in Figure~\ref{contig}, vertex ``GGCA'' is at the left end of contig ``TGCCGTAC'', since its neighbor ``CTGC'' is of type $\langle m$-$n\rangle$. The contig labeling operation first recognizes contig-ends in two supersteps: (1)~in superstep~1, every vertex of type $\langle m$-$n\rangle$ broadcasts its ID to all its neighbors, and then votes to halt; it will never be reactivated again as the remaining computation only involves unambiguous vertices; (2)~in superstep~2, a vertex recognizes itself as a contig-end if it is of type $\langle1\rangle$, or if it is of type $\langle1$-$1\rangle$ and receives the ID of any ambiguous vertex sent from superstep~1.

There are two methods to find all maximal unambiguous paths (i.e., contigs) in $O(\log n)$ supersteps, both of which require contig-end vertices to remove all their edges with ambiguous vertices, so that the DBG graph becomes a set of isolated unambiguous paths, each corresponding to a contig. The first method is to run the simplified S-V algorithm described in Section~\ref{sec:pregel}, so that every vertex $v$ is labeled with the smallest vertex ID in its connected component (i.e., isolated unambiguous path containing $v$). The second method is to use the idea of list ranking described in Section~\ref{sec:pregel} to find all unambiguous paths in $O(\log \ell_{max})$ time, where $\ell_{max}$ is the length of the longest unambiguous path. We now describe the second algorithm in more detail.

We illustrate this algorithm using the example of Figure~\ref{contig}, which is replotted in Figure~\ref{biLR}. Each edge is plotted along with its equivalent edge in the other direction as determined by Property~\ref{prop}, and each vertex is denoted by its integer ID (e.g., ``GGCA'' is encoded with bitmap 10100100, which is 164). As mentioned two paragraphs before, in superstep~2, a vertex $v$ that recognizes itself as a contig-end needs to remove edges with any ambiguous vertex. Instead of deleting such an edge from $v$'s adjacency list, we replace it with a self-loop edge, but we flip the second most significant bit of target ID (i.e., $v$'s ID, let it be $id_v$) to indicate that $v$ is a contig-end. The flipped ID is denoted by $\overline{id_v}$. For example, in Figure~\ref{biLR}, vertex $v_2$ with ID 164 has two neighbors $v_1$ and $v_3$, and it replaces the edge with the ambiguous neighbor $v_1$ (who sent its ID to $v_2$ in superstep~1) by a self-loop edge to $v_2$ itself, leading to a pair of neighbor ID $(\overline{164}, 105)$. Recall from Figure~\ref{id} that the second most significant bit of a 64-bit ID is neither used to encode $k$-mer sequence nor used to differentiate ID types.

In our list ranking approach, each unambiguous vertex maintains a pair of IDs, which is initialized as the pair of neighbor IDs set by superstep~2, as illustrated by round~0 in Figure~\ref{biLR}. Note that a vertex $v$ of type $\langle1\rangle$ also has a pair of IDs, since its {\em NULL} neighbor is replaced with the self-loop edge (note that $v$ is a contig-end). We then perform list ranking in both sequencing directions of a contig, and we call the process as {\em bidirectional list ranking}. We pass messages in both directions rather than from one end of a contig to the other end, since the two ends are symmetrically recognized in superstep~2, and edge direction alone is not sufficient to determine the sequencing direction as explained by Property~\ref{prop}.

In the ID-pair maintained by a vertex $v$, each ID corresponds to $v$'s predecessor in one sequencing direction, which is updated as the predecessor's predecessor after each round until it becomes the flipped ID of a contig-end. We illustrate this process by considering vertex $v_3$ with ID 105 in Figure~\ref{biLR}. In round~0, $v_3$ sends its ID to its two predecessors 164 ($v_2$) and 26 ($v_4$) in one superstep. In the next superstep, $v_2$ receives $v_3$'s ID 105, checks its ID pair $(105, 108)$ and finds the predecessor that is not the received ID, i.e., $v_5$ (108), which is responded back to $v_3$. Similarly, $v_3$ will also receive $\overline{164}$ from $v_2$, and it then replaces its current ID-pair with the received pair of IDs $(\overline{164}, 108)$. In round~1, $v_3$ send requests to 108 ($v_5$) only since it has already reached the contig-end $\overline{164}$ in the other direction. It receives $v_5$'s predecessor $\overline{177}$, and updates its ID-pair as $(\overline{164}, \overline{177})$. Since it reaches both contig-ends, it votes to halt and will not participate in any future computation. In fact, all vertices reach both contig ends before round~2 and vote to halt, and thus the computation stops in 2 rounds. It is not difficult to see that the number of hops between a vertex and its precedessors gets doubled by each round, and thus the computation stops in $O(\log\ell_{max})$ supersteps and is thus a BPPA. When the computation terminates, the ID-pair of each vertex contains the flipped IDs of its two contig-ends. Obviously, each ID-pair uniquely defines a contig, and we use the smaller contig-end vertex's ID as the contig-label.

Bidirectional list ranking alone is not sufficient if the DBG contains a cycle of vertices of type $\langle1$-$1\rangle$ (a special contig if large enough), since these vertices will never reach an end. Note that if the vertex ID-pairs in any contig is not finalized, each round will have some vertices vote to halt due to reaching both contig-ends. Therefore, if the number of active vertices is larger than 0 and does not decrease after a round, the algorithm turns to run our simplified S-V algorithm on the remaining active vertices, so that each vertex in a cycle obtains the smallest ID in the cycle. Bidirectional list ranking is preferred since each round only takes 2 supersteps, much smaller than that required by a round of the S-V algorithm. On the other hand, running the S-V algorithm over vertices in cycles at last is fast, since there are very few active vertices remaining.

To summarize, if only the simplified S-V algorithm is adopted, then each vertex obtains its contig-label as the smallest vertex ID in its contig; if bidirectional list ranking is adopted, each vertex in a non-cycle contig obtains its contig-label as the smaller contig-end vertex's ID, while each vertex in a cycled contig obtains its contig-label as the smallest vertex ID in the cycle.

\vspace{1mm}

\noindent{\bf \textcircled{3} Contig Merging.} This operation takes the labeled unambiguous vertices as the input, and uses a mini MapReduce procedure to group the vertices by their labels. All vertices with the same contig-label are input to {\em reduce}(.), which then merges the sequences of these vertices to obtain the contig.

We now describe the merging process in {\em reduce}(.). Firstly, a hash table is constructed over all the vertices in the contig-group, so that we can lookup a vertex object (storing information like its sequence and neighbors) using its 64-bit integer ID. We also identify a contig-end vertex, which contains a neighbor not in the group (either {\em NULL} or of type $\langle m$-$n\rangle$), to start the stitching with. If such a vertex cannot be found, the contig is cycled and we start stitching from an arbitrary vertex.

We then order all the vertices from the starting vertex (and meanwhile, set the edge directions properly), so that they can be stitched in order. Let us denote the starting vertex by $v_1$, and denote the subsequent vertices after ordering by $v_2, v_3, \ldots, v_k$. Initially, we find a neighbor of $v_1$ that is not its self-loop, which is found as $v_2$. We let $v_1$'s out-neighbor be $v_2$, and let the other neighbor of $v_1$ be its in-neighbor. Edge directions and polarities are properly adjusted using Property~\ref{prop} if they are originally inconsistent. We then obtain $v_2$ from the hash table for processing, using its ID stored in $v_1$'s adjacency list. Generally, for each vertex $v_i\ (i>1)$, we let $v_{i-1}$ be its in-neighbor, and let the other neighbor (which is found as $v_{i+1}$) be the out-neighbor; then $v_{i+1}$ can be obtained from the hash table (using its ID in $v_i$'s adjacency list) to continue the ordering process. The ordering finishes when all $k$ vertices have been processed.

If $v_k$ is of type $\langle1\rangle$, we exit {\em reduce}(.) if the aggregated contig length is not above the user-specified tip-length threshold (since the contig is a tip). In all other cases, we stitch the vertices in the order of $v_1, v_2, \ldots, v_k$ to construct the contig. Specifically, if the edge polarity on $v_1$'s side is $H$, we reverse-complement $v_1$'s sequence and append it to the contig's sequence; otherwise, $v_1$'s sequence is directly appended to the contig's sequence. For each subsequent vertex $v_i\ (i>1)$, we check whether the edge polarity on its side is $L$ or $H$, and use $v_i$'s sequence or its reverse complement to update the contig's sequence. Note that the two sequences overlap by $(k-1)$ elements and this should be taken into consideration. For example, in Figure~\ref{contig}, assume that vertex ``GGCA'' already appended sequence ``TGCC'' to the contig's sequence (as edge polarity is $H$ on the side of ``GGCA''), then vertex ``CGGC'' should only append the complement of the first element (not the last element as as edge polarity is $H$ on the side of ``CGGC'') to the contig's sequence, leading to ``TGCCG''. We also set the contig's coverage as the minimum edge coverage seen during the concatenation, and set the contig's two neighbors with $v_1$'s in-neighbor and $v_k$'s out-neighbor.

\vspace{1mm}

\noindent{\bf \textcircled{4} Bubble Filtering.} The contigs previously constructed may then enter the ``bubble filtering'' operation for further filtering though a mini MapReduce procedure. In {\em map}(.), each contig with neighbors $nb_1$ and $nb_2$ ($nb_1<nb_2$), both of type $\langle m$-$n\rangle$, associates itself with a key $(nb_1, nb_2)$ for shuffling. As a result, all contigs that share two neighboring ambiguous vertices $(nb_1, nb_2)$ are input to {\em reduce}(.), and let us denote them by $c_1, c_2, \ldots, c_k$. We then process each contig $c_i$ as follows: if $c_i$ is not already pruned, we check whether any contig $c_j$ ($j>i$) can prune $c_i$. Specifically, we first compute the edit distance between $c_i$'s sequence and $c_j$'s sequence or its reverse complement (depending on whether $c_i$ and $c_j$'s edge directions are consistent, i.e., $nb_1$-to-$nb_2$ or $nb_2$-to-$nb_1$). If the distance is smaller than a user-defined threshold, we mark $c_i$ (resp.\ $c_j$) as pruned if its coverage is smaller than $c_j$ (resp.\ $c_i$).

\vspace{1mm}

\noindent{\bf \textcircled{5} Tip Removing.} This operation takes both ambiguous $k$-mers and the merged contigs as input. We first need to update the adjacency lists of the ambiguous $k$-mers, to link them to the newly merged contigs. In fact, since some contigs may have been removed due to bubble filtering, some ambiguous $k$-mers may have changed their types from $\langle m$-$n\rangle$ to $\langle 1$-$1\rangle$ or $\langle1\rangle$. Recall from Section~\ref{ssec:ds} that a $k$-mer vertex stores its contig neigbhor by maintaining (1)~the contig vertex's ID (e.g., for requesting its sequence), (2)~the vertex that the contig connects to on the other end, and (3)~other contig information like its length. We set the adjacency lists of the $k$-mer vertices in two supersteps: (i)~in superstep~1, each contig vertex sends its information mentioned above to both neighbors (if not {\em NULL}); then (ii)~in superstep~2, each $k$-mer vertex collects these information into its adjacency list.

Since only path length is concerned during tip removing, we only need to check the $k$-mer vertices as each $k$-mer vertex $u$ maintains each contig neighbor $c$ in the form of $c$'s ID, $c$'s sequence length, and the $k$-mer vertex $v$ on the other end of $c$. However, when deleting the edge $(u, v)$ (due to being part of a tip), a message should be sent to the contig vertex $c$ (using $c$'s ID stored in the adjacency list item) to tell it to delete itself, which we take for granted and will not emphasize in the subsequent algorithm description.

Note that the removal of tips may cause some vertices of type $\langle m$-$n\rangle$ to change their type to $\langle1\rangle$, hence generating new tips. As a result, we run our vertex-centric tip removing procedure for multiple phases, until no new $\langle1\rangle$-typed vertex is generated at the end of a phase.

In a phase, we start message passing from vertices of type $\langle1\rangle$, where a message records (1)~the sender's ID, (2)~cumulative sequence length, and (3)~a type {\em REQUEST}. A $\langle1\rangle$-typed vertex $u$ initializes the cumulative sequence length as $k$ (i.e., $u$'s sequence length). When a vertex $u$ of type $\langle 1$-$1\rangle$ receives a {\em REQUEST} message, it relays the message to the other neighbor $v$ (i.e., not the sender) by adding the cumulative sequence length by 1 (contributed by $u$) plus the contig length minus $(k-1)$ if edge $(u, v)$ contains a contig (``minus $(k-1)$'' is not to count the overlapping sequence). 

The {\em REQUEST} message ends at an $\langle m$-$n\rangle$-typed or $\langle1\rangle$-typed vertex $v$, which checks whether the cumulative sequence length is not larger than the tip-length threshold. If so, $v$ sends a message of type {\em DELETE} to the sender to delete the vertices on the dangling path. The {\em DELETE} message is relayed by $\langle 1$-$1\rangle$-typed vertices back till reaching the $\langle1\rangle$-typed vertex that initiates the {\em REQUEST} message, and vertex and contig deletions are triggered along the backward message propagation.

A special case is when a tip has two $\langle1\rangle$-typed ends. Since both vertices at the ends initiate a {\em REQUEST} message sent towards each other, when the two {\em DELETE} messages are sent back, they meet in the middle of the tip (rather than reach the other $\langle1\rangle$-typed end).

An $\langle m$-$n\rangle$-typed vertex $v$ also deletes its edge to the neighbor that it sends a {\em DELETE} message, and if its type becomes $\langle1\rangle$, it keeps itself activated to initiate the {\em REQUEST} message in the next phase.

\section{Experiments}\label{sec:results}
The state-of-the-art parallel assemblers often use ad-hoc design. For example, ABySS~\cite{abyss} builds the DBG by letting each $k$-mer send messages to its 8 possible neighbors (with A/T/G/C prepended/appended) to establish edges. This increases ambiguity (and hence reduces contig length) since an edge will be created between 2-mers ``CA'' (e.g., contributed by 3-mer ``CAT'') and ``AA'' (e.g., contributed by ``GAA'') even though the 3-mer ``CAA'' does not exist in the DNA molecule. As another example, instead of using list ranking or S-V to find maximal unambiguous paths as contigs, Spaler~\cite{spaler} iteratively breaks each unambiguous path by sampled vertices to form segments, and then merges segments that meet at a sampled boundary vertex. The process is repeated until $\langle m$-$n\rangle$-typed vertices account for more than 1/3 of all vertices in the graph, and this heuristic provide no guarantee of path maximality. PPA-assembler avoids ad-hoc design by well-designed assembly algorithmic logic (that can be further customized by users), which are decoupled from low-level distributed communication through using the performance-optimized vertex-centric and mini MapReduce procedures provided by Pregel+.

We compare PPA-assembler with the state-of-the-art parallel assemblers, ABySS (version~1.5.2), Ray (version~2.3.1) and SWAP-Assembler (version~3.0). Spaler is not open-sourced and is thus not included in our comparison. We use the simple workflow \textcircled{1}\textcircled{2}\textcircled{3}\textcircled{4}\textcircled{5}\textcircled{6}\textcircled{2}\textcircled{3} in Figure~\ref{overview} for PPA-assembler, i.e., to grow contigs once further after error correction. However, we remark that users may customize their own workflow or even change the existing operations (e.g., add coverage-threshold pruning to bubble filtering) or add new operations implemented in Pregel+'s API (e.g., branch splitting~\cite{spaler} for error correction) to implement different assembly strategies, including those of ABySS, Ray and SWAP-Assembler.

All experiments were conducted on a cluster of 16 machines connected by Gigabit Ethernet, each with 48 GB DDR3 RAM and 12 cores (two Intel Xeon E5-2620 CPUs). We used $k=31$ for defining $k$-mers. For PPA-assembler, edit distance threshold for bubble filtering is set as 5, and length threshold for tip removing is set as 80, since we found that the sequencing results are very stable near these parameter ranges. We used the default settings of ABySS, Ray and SWAP-Assembler in their respective experiments.

\begin{table}[t]
	\centering
	\caption{Datasets (M = 1,000,000, bp = base pairs)}\label{data}
	\vspace{-2mm}
	\includegraphics[width=\columnwidth]{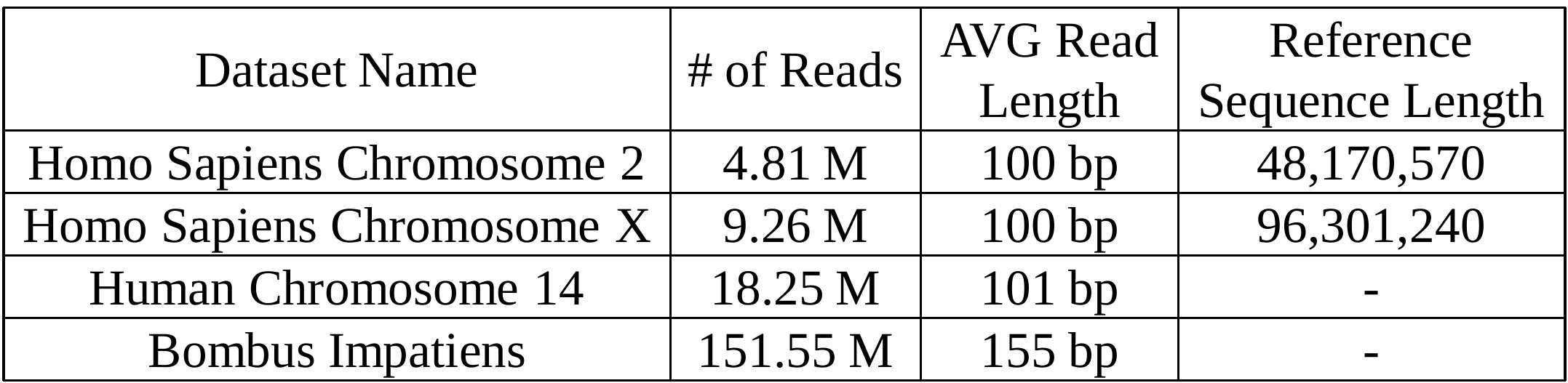}
	\vspace{-8mm}
\end{table}

\begin{figure*}[t]
	\centering
	\includegraphics[width=1.8\columnwidth]{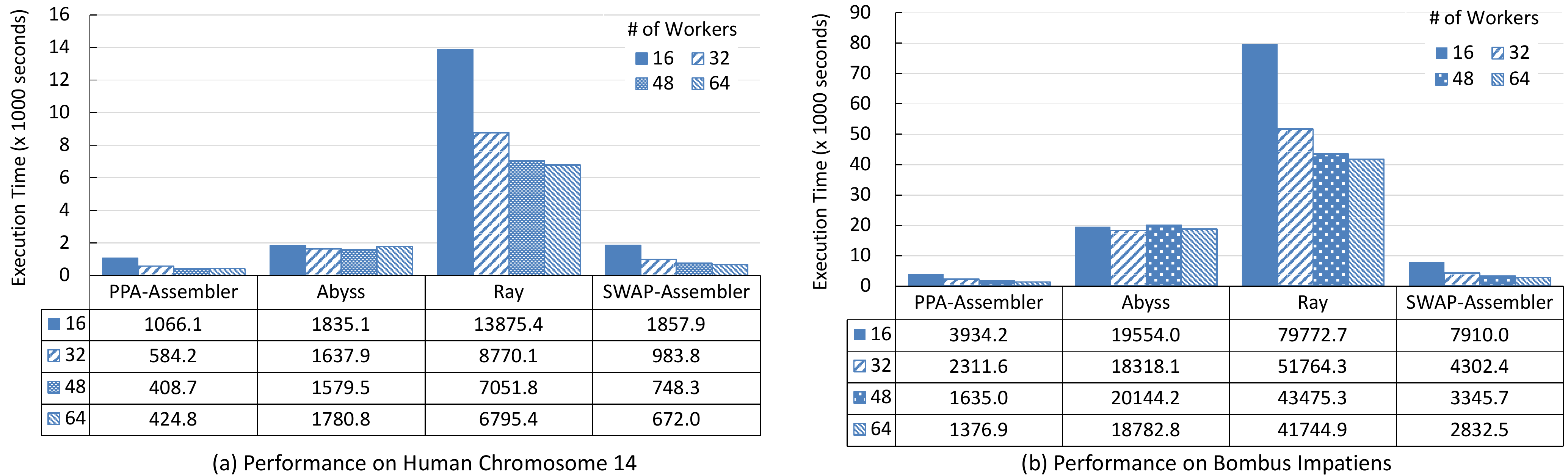}
	\vspace{-2mm}
    	\caption{Execution Time of the Assemblers with Varying Number of Machines}\label{scale}
   	 \vspace{-4mm}
\end{figure*}

We ran our experiments with the 4 datasets shown in Table~\ref{data}, where are listed in the increasing order of data volume. The two smaller datasets are generated from NCBI's reference gene sequences Homo Sapiens Chromosome X (HC-X)\footnote{\scriptsize\url{http://www.ncbi.nlm.nih.gov/nuccore/NC_000023.11}} and Homo Sapiens Chromosome 2 (HC-2)\footnote{\scriptsize\url{https://www.ncbi.nlm.nih.gov/nuccore/NC_000002.12}}, using the ART\footnote{\scriptsize\url{http://www.niehs.nih.gov/research/resources/software/biostatistics/art/}} software~\cite{huang2012art}. These datasets have reference sequences and thus we can measure the sequencing quality exactly. To test scalability further, we also selected the 2 larger datasets Human Chromosome 14 (HC-14)~\footnote{\scriptsize\url{http://gage.cbcb.umd.edu/data/Hg_chr14}} and Bombus Impatiens (BI)~\footnote{\scriptsize\url{http://gage.cbcb.umd.edu/data/Bombus_impatiens}}, which are downloaded from the GAGE project~\cite{salzberg2012gage}. All the datasets are in FASTQ format, which includes the sequence of each DNA read.

\vspace{1mm}

\noindent{\bf Running Time \& Scalability.} We use bidirectional list ranking for contig labeling in this set of experiments. Figure~\ref{scale} shows the performance of the four assemblers on the two large datasets HC-14 and BI (the results on HC-X and HC-2 are similar and are thus omitted). For each assembler, we show the end-to-end execution time of assembly when each machine runs 1, 2, 3 and 4 workers, respectively. We can see that PPA-assembler is much faster than all the other assemblers in all cases, thanks to the efficient Pregel+ backend and our well-designed algorithms. This performance difference becomes even larger when the input dataset becomes larger (e.g., compare Figure~\ref{scale}(b) to Figure~\ref{scale}(a)), which verifies the superior scalability of PPA-assembler towards data size. In contrast, Ray has the poorest performance, often one order of magnitude slower than the other assemblers.

As for the scalability towards the number of workers, the performance of PPA-assembler, SWAP-Assembler and Ray keeps improving as the worker number increases. In contrast, the performance of ABySS is insensitive to the number of workers. In fact, more workers may even lead to a longer assembly time.

\vspace{1mm}

\begin{table}[t]
	\centering
	\caption{LR v.s.\ S-V for Labeling Unambiguous $k$-mers}\label{result5}
	\vspace{-2mm}
	\scalebox{0.94}{
		\begin{tabular}{|c|c|c|c|c|c|c|} \hline
			\multirow{2}{*}{Datasets} & \multicolumn{2}{c|}{\# of Supersteps} & \multicolumn{2}{c|}{\# of Messages} & \multicolumn{2}{c|}{Runtime (s)}\\ \cline{2-7}
			& LR & S-V & LR & S-V & LR & S-V  \\ \hline
			HC-X & 26 & 86 & 2,325 M & 5,913 M & 93 & 212 \\ \hline
			HC-2 & 28 & 93 & 1,498 M & 3,644 M & 58 & 128 \\ \hline
			HC-14 & 67 & 93 & 2,342 M & 6,852 M & 213 & 415 \\ \hline
			BI & 60 & 86 & 6,705 M & 22,958 M & 239 & 723 \\
			\hline\end{tabular}
	}
	\vspace{-2mm}
\end{table}

\begin{table}[t]
	\centering
	\caption{LR v.s.\ S-V for Labeling Contigs}\label{result6}
	\vspace{-2mm}
	\scalebox{0.93}{
		\begin{tabular}{|c|c|c|c|c|c|c|} \hline
			\multirow{2}{*}{Datasets} & \multicolumn{2}{c|}{\# of Supersteps} & \multicolumn{2}{c|}{\# of Messages} & \multicolumn{2}{c|}{Runtime (s)}\\ \cline{2-7}
			& LR & S-V & LR & S-V & LR & S-V  \\ \hline
			HC-X & 32 & 44 & 2.16 M & 5.28 M & 0.51 & 0.67 \\ \hline
			HC-2 & 12 & 37 & 1.05 M & 2.74 M & 0.20 & 0.50 \\ \hline
			HC-14 & 22 & 51 & 6.04 M & 22.46 M & 1.06 & 1.83 \\ \hline
			BI & 38 & 65 & 74.36 M & 280.04 M & 3.77 & 10.26 \\
			\hline\end{tabular}
	}
	\vspace{-6mm}
\end{table}

\noindent{\bf Bidirectional List Ranking v.s.\ Simplified S-V.} These are two approaches for contig labeling. As we mentioned, while both algorithms are PPAs that runs for $O(\log n)$ rounds (and hence supersteps), each round of S-V require a larger number of supersteps than a round in list ranking, and thus list ranking (LR) is expected to be much faster. Recall that we run PPA-assembler with the simple workflow of \textcircled{1}\textcircled{2}\textcircled{3}\textcircled{4}\textcircled{5}\textcircled{6}\textcircled{2}\textcircled{3} in Figure~\ref{overview}, and ``\textcircled{2} contig labeling'' is performed twice: once for labeling unambiguous $k$-mers, and once for labeling contigs (to grow longer ones).

Table~\ref{result5} and Table~\ref{result6} show the comparison of LR and S-V for labeling $k$-mers and labeling contigs, respectively, on the four datasets, where we report (1)~the number of supersteps, (2)~the number of messages, and (3)~the running time. We can see that LR runs for much fewer supersteps, sends much fewer messages, and is much faster than S-V. The message number and runtime in Table~\ref{result6} is three orders of magnitude less than those in Table~\ref{result5}, since the vertex number is significantly reduced after we merge unambiguous $k$-mers into contigs. For example, the DBG of the HC-2 dataset has 46.97 M vertices, which is reduced to 1.00 M vertices after merging unambiguous $k$-mers into contigs, and further to 68,264 vertices after these contigs are merged after error correction.

\vspace{1mm}

\noindent{\bf Sequencing Quality.} We now assess the sequencing quality of the assemblers. We remark that, for PPA-assembler, we are just evaluating the adopted workflow. We can easily configure PPA-assembler with other assembly strategies that leads to a higher sequencing quality. Even with the adopted workflow, PPA-assembler achieves comparable (if not better) quality, which we present next.

We used the popular assessment tool, QUAST~\cite{gurevich2013quast}, which reports various quality metrics commonly used in genetic analysis. These metrics include: (1)~N50, which is defined as the sequence length of the contig that contains middle element of the sequence that concatenates all contigs from the longest one to the shortest one; (2)~the number of contigs whose length are larger than 500 bp; (3)~the length of largest contig; (4)~the total length of contigs; (5)~genome coverage, which is the percentage of bases in the genome covered; (6)~the number of misassembled contigs; (7)~unaligned length, and so on. Some of these metrics do not require a reference sequence (e.g., N50), while others do (e.g., the number of misassembled contigs).

\begin{table}[t]
	\centering
	\caption{Quality Comparison on HC-2}\label{result2}
	\vspace{-2mm}
	\scalebox{0.85}{
		\begin{tabular}{|l|c|c|c|c|} \hline
			Assembler & PPA & ABySS & Ray & SWAP \\ \hline
			\# of contigs & 22,707 & {\color{red}{29,231$^*$}} & 26,739 & 12,477 \\ \hline
			Total length & {\color{red}{36,878,742$^*$}} & 31,426,810 & 20,854,349 & 8,232,160 \\ \hline
			N50 & {\color{red}{2,070$^*$}} & 1,184 & 779 & 640 \\ \hline
			Largest contig & {\color{red}{16,376$^*$}} & 7,166 & 3,248 & 1,982 \\ \hline
			GC (\%) & 40.89 & {\color{red}{41.77$^*$}} & 41.03 & 41.21 \\ \hline
			\# Misassemblies & {\color{red}{1$^*$}} & 4 & {\color{red}{1$^*$}} & 167 \\ \hline
			Misassembled length & 1,366 & 3,666 & {\color{red}{520$^*$}} & 115,998 \\ \hline
			Unaligned length & {\color{red}{24$^*$}} & 427 & 1,227 & 47,810 \\ \hline
			Genome fraction (\%) & {\color{red}{76.285$^*$}} & 65.104 & 42.981 & 16.963 \\ \hline
			\# Mismatches per 100 kbp & {\color{red}{0.43$^*$}} & 13.75 & 1.04 & 43.02 \\ \hline
			\# Indels per 100 kbp & {\color{red}{0.03$^*$}} & 0.10 & 0.09 & 5.32 \\ \hline
			Largest alignment & {\color{red}{16,376$^*$}} & 7,166 & 3,248 & 1,982 \\
			\hline\end{tabular}
	}
	\vspace{-6mm}
\end{table}

We present the sequencing quality of the assemblers on HC-2 in Table~\ref{result2}. Since HC-2 has a reference sequence, we obtained all the various quality metrics reported by QUAST. The best results among the assemblers are highlighted in Table~\ref{result2}, and we can observe that PPA-assembler performs the best in the majority of the metrics (and comparable in others). For example, it has the highest N50 value and the lowest misassemblies (i.e., only 1 misassembled contig). It is worth mentioning that the second round of contig merging is effective: N50 is 1074 after we merge unambiguous $k$-mers into contigs, and it improves to 2070 (i.e., is doubled) after we merge contigs after error correction. The results on HC-X are similar and thus omitted.

\begin{table}[h]
	\centering
	\caption{Quality Comparison on HC-14}\label{result3}
	\vspace{-2mm}
	\scalebox{0.85}{
		\begin{tabular}{|l|c|c|c|c|} \hline
			Assembler & PPA & ABySS & Ray & SWAP \\ \hline
			Number of contigs & 41,445 & 18,008 & 45,984 & {\color{red}{47,252$^*$}} \\ \hline
			Total length & 62,667,868 & 26,586,604 & 63,456,459 & {\color{red}{63,752,569$^*$}} \\ \hline
			N50 & {\color{red}{1,891$^*$}} & 1,847 & 1,641 & 1,605 \\ \hline
			Largest contig & {\color{red}{16,069$^*$}} & 15,744 & 15,116 & 13,251 \\
			\hline\end{tabular}
	}
	\vspace{-4mm}
\end{table}

We report the sequencing quality of the assemblers on HC-14 in Table~\ref{result3}. Since HC-14 has no reference sequence, we cannot report many of the metrics such as misassemblies. The results show that PPA-assembler achieves the largest N50 value, performs the best in 2 of the 4 metrics, and achieves comparable performance to the best in the other two metrics. The results on BI are similar and thus omitted.

\section{conclusion}\label{sec:conclude}
We presented a scalable and flexible de novo genome assembler, PPA-assembler, built on a popular big data framework and provides strict performance guarantee. PPA-assembler is many times faster than other distributed assemblers, and achieves comparable (if not better) sequencing quality.

{\small
\bibliographystyle{abbrv}
\bibliography{ref_dna}
}

\end{document}